\documentclass{imsart}

\pubyear{0000}
\volume{00}
\issue{0}
\doi{0000}
\firstpage{1}
\lastpage{1}

\usepackage{imsart,natbib}
\RequirePackage[colorlinks,citecolor=blue,urlcolor=blue]{hyperref}
\usepackage{graphicx,comment}
\usepackage{booktabs,longtable}
\usepackage{bayes}
\def\bas#1\eas{\begin{align*}#1\end{align*}}

\usepackage{amsmath,bm,amsfonts,amsthm}
\usepackage[usenames]{color}
\usepackage{graphicx}
\usepackage{booktabs,longtable}
\usepackage{lineno}

\newcommand{\bb}{\bm{\beta}}

\newcommand{\commentt}[1]{}

\newcommand{\bX}{\boldsymbol{X}}

\newcommand{\lam}{\boldsymbol{\Lambda}}

\newcommand{\bz}{\boldsymbol{z}}

\let\originalleft\left
\let\originalright\right
\renewcommand{\left}{\mathopen{}\mathclose\bgroup\originalleft}
\renewcommand{\right}{\aftergroup\egroup\originalright}

\DeclareMathOperator{\pop}{pop}
\DeclareMathOperator{\distinct}{distinct}

\definecolor{darkred}{RGB}{170,0,0}
\definecolor{darkgreen}{RGB}{0,153,0}

\begin{document}



\begin{frontmatter}


\title{{Entity Resolution with \\ Empirically Motivated Priors}}

\runtitle{{Entity Resolution with Empirically Motivated Priors}}


\begin{aug}

\author{\fnms{Rebecca C. }\snm{Steorts}\thanksref{addr1}
\ead[label=e1]{beka@cmu.edu}
}

\runauthor{R. C. Steorts}

\address[addr1]{Visiting Assistant Professor, Department of Statistics, Carnegie Mellon University, 5000 Forbes Avenue, Pittsburgh, PA 15213, \printead{e1}}

\end{aug}

\begin{abstract}
Databases often contain corrupted, degraded, and noisy data with duplicate entries across and within each database. Such problems arise in citations, medical databases, genetics, human rights databases, and a variety of other applied settings. The target of statistical inference can be viewed as an unsupervised problem of determining the edges of a bipartite graph that links the observed records to unobserved latent entities.
Bayesian approaches provide attractive benefits, naturally providing  uncertainty quantification via posterior probabilities. 
We propose a novel
record linkage approach based on empirical Bayesian principles.  Specifically, the empirical Bayesian--type step consists of taking the empirical distribution function of the data as the prior for the latent entities.
This approach improves on the earlier HB approach not only by avoiding the prior specification problem but also by allowing both categorical and string-valued variables. 
Our extension to string-valued variables also involves the proposal of a new probabilistic mechanism by which observed record values for string fields can deviate from the values of their associated latent entities.  Categorical fields that deviate from their corresponding true value are simply drawn from the empirical distribution function.
We apply our proposed methodology to a simulated data set of German names and an Italian household survey, showing our method performs favorably compared to several standard methods in the literature.
We also consider the robustness of our methods
to changes in the hyper-parameters. 
\end{abstract}

\end{frontmatter}


\newcounter{itemnum}
\newtheorem{lemma}{Lemma}
\theoremstyle{definition}
\newtheorem*{rem}{Remark}

\section{Introduction} 
\label{sec:intro}
Entity resolution, also known as record linkage, de-duplication, or co-reference resolution \citep{christen_2011},
is the merger of multiple databases and/or removal of duplicated records within a database in the absence of unique record identifiers.
Traditional entity resolution methods are based upon simple, unsupervised approaches to find links between co-referent records \citep{fellegi_1969}.  These approaches compute pairwise probabilities of matching for all pairs of records, which is computationally infeasible for databases of even moderate size \citep{winkler_2006}. An alternative to record-to-record comparisons is the clustering of records to an unobserved latent entity.
Such a clustering structure can be conceptualized as a bipartite graph with edges linking an observed record to the latent entity to which it corresponds. Each latent entity has a ``true'' value for each field included in the database, and the field values of the associated records can be distorted from the ``true'' value with some probability. This methodology was introduced by \citet{steorts_2013b, steorts_2014_aistats} with a hierarchical Bayesian (HB) model, in which records are clustered to latent entities and the values of the latent entities are assigned prior distributions through a high-dimensional data structure.  (For brevity, we will refer to \citet{steorts_2013b}, but for more details see \citet{steorts_2014_aistats}.)
This contribution unified the processes of record linkage and de-duplication under a single framework.
Nevertheless, the approach of \cite{steorts_2013b} was limited in some respects.  First, it could only be applied to categorical data. 
In practice, record linkage problems often include string-valued data such as names, addresses, etc.  The treatment of such variables as categorical typically results in poor performance since it ignores the notion of distance between strings that do not exactly agree.
Second, the hierarchical Bayesian model required the specification of priors for the latent entity values, which can be quite difficult in many applied settings.

We propose methodology, clustering records to a hypothesized latent entity,
{with the empirical distribution of the data for each field used as the prior for the corresponding field values of the latent entities.  Our model handles both categorical and noisy ``text'' data.}
We seek to develop unsupervised learning approaches for entity resolution in
the absence of
high-quality training data, which is
{often the case}
in many real-world applications such as online medical records, genetics data,
records of
human rights violations, and official statistics. In our approach, we advocate an EB formulation, in which the prior for the latent entity value for each field is taken as the empirical distribution of the data values for that field.  This EB approach both simplifies the model and eliminates the need to specify subjective priors for the latent entity values. 
 Moreover, the simplification of the model eases the computational burden imposed by the required MCMC procedures.  Our second major improvement to the record linkage literature is that we allow the records to include both categorical and string-valued variables.  For string-valued variables, we model the distortion (i.e., the departures of the record values from their associated latent individual values) using a probabilistic mechanism based on some measure of distance between the true and distorted strings.  Our approach is flexible enough to permit the use of a variety of string distances, which can thus be chosen to suit the needs of any given application. We apply our proposed methodology to two datasets: a simulated dataset of German names and a data set from the Italian Survey on Household and Wealth.  For both datasets, we show that our method compares favorably to existing approaches in the literature.
Furthermore, we illustrate the robustness of our methods on both datasets in terms of the hyper parameters/unknown parameters. 

\subsection{Prior Work}
A variety of techniques for record linkage have been proposed, originally by
\citet{fellegi_1969}, who gave the first mathematical model for one-to-one entity resolution across two databases. \citet{sadinle_multi_1} extended this approach to linking records across $k>2$ databases.
Their approach is computationally infeasible for large-scale record
linkage, since it requires the estimation of $2^N-1$ 
conditional probabilities for databases with $N$ records.
More sophisticated approaches have typically employed supervised or semi-supervised learning techniques in the disambiguation literature \citep{han_2004, torvik_2009, giles_2009, martins}.
However, such methods assume the existence of large, accurate sets of training data, which are often difficult and/or expensive to obtain.   We develop
unsupervised
learning approaches for de-duplication for applications that lack high-quality training data. One popular method that we compare to is that of random forests \citep{breiman_2001}, which are ensembles of classification trees trained on bootstrap samples of the training data.  Random forests provide a powerful method of aggregating classification trees to improve prediction in the decision tree framework. The predicted class from the random forest is the class that receives the majority of the class votes of the individual trees. In our context, the covariates of the trees are similarity scores, the training data are the pairwise comparisons of the labeled records, and the binary-valued response class is simply match/non-match.  A tree's class prediction for any pair of records
assigns the majority class vote (match vs.\ non-match) for the pair's terminal node. Such methods have been extended and used by \citet{ventura_2013} for author disambiguation. 
Another approach is provided by Bayesian Adaptive Regression Trees (BART) \citep{george_2010} applied to the same setup of covariates and responses.
\citet{winkler_2006} provides an overview of both supervised and
unsupervised entity resolution techniques.

Other related work appears in the statistics, computer science, and machine learning literature, where the common theme is typically clustering or latent variable models.
One common application of interest is the disambiguation of document authors.
\citet{getoor_2006} describe an entity-resolution approach  based on latent Dirichlet allocation, which infers the total number of unobserved entities (authors). A requirement of this approach is that the number of co-authorship groups must be known/estimated. Furthermore, labeled data is required for setting parameters in their model. 
In the work of \citet{dai_2011},  groups of authors are associated with topics instead of individual authors, using a non-parametric Dirichlet process. However, when clustering records to latent topics, the number of latent topics typically does not grow as fast as the number of records. It is well known that if the number of data points (records) grows and the number of latent clusters (entities) grows more slowly or remains fixed, then the latent clusters are not exchangeable. Hence, the Dirichlet mixture model, the Pitman-Yor process, and other related models (Kingman paintbox) are inappropriate \citep{broderick_2014_vb, wallach_2010}. 

Bayesian methods have a long history of use in record linkage models. A major advantage of Bayesian methods is their natural handling of uncertainty quantification for the resulting estimates.
Within the Bayesian paradigm,
most work has focused on specialized approaches related to linking two files \citep{gutman_2013, liseo_2011, larsen_2001, belin_1995}.  These contributions, while valuable, do not easily generalize to more than two files or to de-duplication. For a review of recent development in Bayesian methods, see \citet{liseo_2013}. De-duplication for more than two files was explored by \citet{sadinle_multi_1}.  These methods were found to be computationally infeasible for large databases as the order of the algorithm was $O(N^k)$, where $N$ is the total number of records
and $k$ is the number of files. 

Recent advances were made by \citet{steorts_2013b}, who introduced a hierarchical Bayesian (HB) model that simultaneously handled record linkage and de-duplication for categorical data.
Their approach allowed for natural uncertainty quantification during analysis and post-processing.  Also, they developed a framework for reporting a point estimate of the linkage structure. Further advancements were made
by
\cite{sadinle_2014}, who
extended
to string variables and
used
a ``coreference matrix" as a prior on partitions of the linkages. This work
has
the same features as our proposed work in taking advantage of the Bayesian paradigm: 
it
allows the incorporation of prior information on the reliability of the field attributes, is unsupervised, and accounts for linkage uncertainty.  \cite{steorts_2014_aistats}
pointed
out the connection between the linkage structure and the coreference matrix.
{However, the likelihood of \citeauthor{sadinle_2014}'s model incorporates the record data only through pairwise similarity scores, whereas our method directly models the actual field data of the records.}

{It should also be noted that there are certain types of seemingly relevant methodology that may in fact be irreconcilable with the basic structure of record linkage.  In particular, it may be asked whether nonparametric techniques can be brought to bear on the record linkage problem.  Unfortunately, such approaches typically entail notions of exchangeability that are inappropriate in the context of record linkage.
\citep[See][for a more thorough discussion.]{broderick_2014_vb}}
\section{Empirical Bayesian Model for Entity Resolution} 
\label{sec-bayes-research}
We use a Bayesian model in the spirit of \citep{steorts_2013b}, but with three major modifications. 
We compare and contrast the two
models
in Appendix \ref{sec:appendix}.
Before introducing {our model,} we first give our notation. 


\subsection{Notation}
\label{sec:notation}
Suppose we have $k$ lists, which we index with~$i$.  The $i$th list has $n_i$ records, which we index with~$j$.  Each record corresponds to one of
a population of $N_{\pop}$
latent individuals, which we index with $j'$.
Note that the number of latent individuals represented by records in the lists is at most $N=\sum_{i=1}^k n_i$, but $N_{\pop}$ may be larger or smaller than~$N$.
Each record or latent individual has values on $p$~fields, which we index with~$\ell.$ (The model of \citet{steorts_2013b} assumed all fields to be categorical, however we do not make this limiting assumption.) 
The number of possible categorical values for the $\ell$th field is $M_\ell$.
 
Next, let $X_{ij\ell}$ denote the
observed value of the $\ell$th field for the $j$th record in the $i$th list,
and let $Y_{j'\ell}$ denote the true value of the $\ell$th field for the $j'$th latent
individual. Let $\lambda_{ij}$ denote the latent individual to which the
$j$th record in the $i$th list corresponds, i.e., $X_{ij\ell}$ and $Y_{j'\ell}$
represent the same individual if and only if $\lambda_{ij}=j'$.
Let $\bm\Lambda$ denote the $\lambda_{ij}$ collectively.
Let $z_{ij\ell}$ be the indicator of whether a distortion has occurred for record field value $X_{ij\ell}$.  Note that if $z_{ij\ell}=0$, then $X_{ij\ell}=Y_{\lambda_{ij}\ell}$.  If instead $z_{ij\ell}=1$, then $X_{ij\ell}$ may differ from $Y_{ij\ell}$.
Let
$\delta_a$ denote the distribution of a point mass at $a$ (e.g., $\delta_{y_{\lambda_{ij}\ell}}$).

\subsection{{Model} for Entity Resolution}
Assume fields $1,\ldots,p_s$ are string-valued, while fields $p_s+1,\ldots,p_s+p_c$ are categorical, where $p_s+p_c=p$ is the total number of fields.

One major novelty addresses the prior distributions of the latent field values $Y_{j'\ell}$ of the latent individuals.  The model of \cite{steorts_2013b} used a HB construction for these priors.  However, such a prior can be extremely difficult to specify subjectively in practice, particularly for string-valued variables. 
Thus, we instead propose an empirical Bayesian approach in which we take the prior distribution of $Y_{j'\ell}$ to be the empirical distribution of the values for field~$\ell$ in the combined set of record data.
For each $\ell\in\{1,\ldots,p_s+p_c\}$, let $S_\ell$ denote the set of \emph{all} values for the $\ell$th field
that occur anywhere in the data, i.e.,
$S_\ell=\{X_{ij\ell}:1\le i\le k, 1\le j\le n_i\}$,
and let $\alpha_\ell(w)$ equal the empirical frequency of value~$w$ in field~$\ell.$
Then let $G_\ell$ denote the empirical distribution of the data in the $\ell$th field from all records in all lists combined.  So, if a random variable~$W$ has distribution $G_\ell$, then for every $w\in S_\ell$,
$P(W=w)=\alpha_\ell(w)$.
Hence, we take $G_\ell$ to be the prior for each latent individual $Y_{j'\ell}$.
We use the frequency of occurrence to 
increase the weight of more ``popular'' entries.
{This approach provides dramatic computational savings in comparison to a hierarchical specification of \citet{steorts_2013b}, especially when considering string-valued fields.  Note that under this approach, the number of possible values for any particular field of a latent entity is no greater than the number of records.  Thus, it is computationally feasible to consider a discrete distribution on this set.  Moreover, certain key quantities that may be necessary for subsequent calculations, such as the string distance between two such values, can be computed a single time in advance for all possible pairs.  In contrast, under a hierarchical specification, a string-valued field of a latent entity could presumably take any value in the set of all strings (up to some maximum length).  Such a set is so large that it presents computational difficulties if it is to serve as the support of a distribution.}

Unlike \citet{steorts_2013b}, in our proposed model, we allow the distortion probability to depend on the list as well as the field, i.e., we take $\beta_{i\ell}$ instead of $\beta_\ell$.  This change reflects the fact that different lists may be compiled using different data collection methods, which may be more or less prone to error.

The aforementioned alterations to the model also necessitate a modification of the distortion model.  If a distortion occurs for a categorical field~$\ell$, we take the distribution of the distorted value to be~$G_\ell$.  If a distortion occurs for a string-valued field~$\ell$, then the probability that the distorted value takes the value~$w$ is given by
\[
P(X_{ij\ell}=w\mid\lambda_{ij},Y_{\lambda_{ij}\ell},z_{ij\ell})
=\frac{\alpha_\ell(w)\,\exp[-c\,d(w,Y_{\lambda_{ij}\ell})]}{\sum_{w\in S_\ell}\alpha_\ell(w)\,\exp[-c\,d(w,Y_{\lambda_{ij}\ell})]},
\]
where $c>0$ is known and $d(\cdot,\cdot)$ is some string distance, or equivalently, one minus some string similarity score.  For brevity, denote this distribution by $F_\ell(Y_{\lambda_{ij}\ell})$.
Our proposed model is 
\par\noindent
\begin{align}
X_{ij\ell}\mid \lambda_{ij},\,Y_{\lambda_{ij}\ell},\,z_{ij\ell}\;&\stackrel{\text{ind}}{\sim}\begin{cases}\delta(Y_{\lambda_{ij}\ell})&\text{ if }z_{ij\ell}=0\\F_\ell(Y_{\lambda_{ij}\ell})&\text{ if }z_{ij\ell}=1\text{ and }\ell\le p_s\\G_\ell&\text{ if }z_{ij\ell}=1\text{ and }\ell>p_s\end{cases} \notag \\
Y_{j'\ell}\;&\stackrel{\text{ind}}{\sim}G_\ell \notag \\
z_{ij\ell}\mid\beta_{i\ell}\;&\stackrel{\text{ind}}{\sim}\text{Bernoulli}(\beta_{i\ell})\notag \\
\beta_{i\ell}\;&\stackrel{\text{ind}}{\sim}\text{Beta}(a,b) \notag \\
\lambda_{ij}\;&\stackrel{\text{ind}}{\sim}\text{DiscreteUniform}\left(1,\ldots,N\right)
\label{model:eb},
\end{align}
where all distributions above are also independent of each other. We assume that $a,b, N$ are assumed known. We explore the sensitivity of these parameters, $c,$ and $d(\cdot)$ in \S \ref{sec:sensitivity}.

\begin{rem}Although each distribution $G_\ell$ is constructed using the observed values of $\bm X$ in the data, this dependency is ignored from the standpoint of computing the posterior under the Bayesian model.  This is merely a standard example of empirical Bayesian methodology.  Although admittedly a bit awkward to interpret from a purely philosophical standpoint, the empirical Bayesian paradigm is quite well attested in both the theory and practice of modern statistics \citep{robbins_1956,carlin_2000}.
\end{rem}

To concisely state the joint posterior of the above model, first define
for each $w_0\in S_\ell$ the quantity
\[
\left[h_\ell(w_0)\right]^{-1}=\sum_{w\in S_\ell}\exp\left[-c\,d(w,w_0)\right].
\]
Note that $h_\ell(w_0)$ can be computed in advance for each possible
$w_0\in S_\ell$. After some simplification, the joint posterior is
\par\noindent
\begin{align*}
&\pi(\bm\lambda,\bm Y,\bm z,\bm\beta\mid\bm X)\\
&\propto
\prod_{i=1}^k\prod_{j=1}^{n_i}\left\{
\left[\mathop{\prod_{\ell=1}^{p_s+p_c}}_{z_{ij\ell}=1}\alpha_\ell(X_{ij\ell})\right]\left[\mathop{\prod_{\ell=1}^{p_s}}_{z_{ij\ell}=1}h_\ell(Y_{\lambda_{ij}\ell})\right]\exp\!\left[-c
\sum_{\ell=1}^{p_s}z_{ij\ell}\,
d(X_{ij\ell},Y_{\lambda_{ij}\ell})\right]
\right\}\\
&\qquad\times\left[\prod_{j'=1}^N\prod_{\ell=1}^{p_s+p_c}\alpha_\ell(Y_{j'\ell})\right]\left[\prod_{i=1}^k\prod_{\ell=1}^{p_s+p_c}\beta_{i\ell}^{\sum_{j=1}^{n_i}z_{ij\ell}+a-1}(1-\beta_{i\ell})^{n_i-\sum_{j=1}^{n_i}z_{ij\ell}+b-1}\right]\\
&\qquad\times I(X_{ij\ell}=Y_{\lambda_{ij}\ell}\text{ for all }i,j,\ell\text{ such that }z_{ij\ell}=0).
\end{align*}
(See Appendix \ref{app:joint} for further details.)
\section{Gibbs Sampler}
\label{sec:gibbs}

Since it is not feasible to sample directly from the joint posterior, inference from the EB model is made via a Gibbs sampler that cycles through drawing from the conditional posterior distributions.  We now provide these conditional distributions explicitly.  Note that notation throughout this section may suppress dependency on variables and/or indices as needed for convenience.

First, consider $\bm\beta \mid \bm{\Lambda}, \bm Y, \bm z, \bm X$.
Let $Z_{i\ell}=\sum_{j=1}^{n_i}z_{ij\ell}$.
Then it is straightforward to show that
\par\noindent
$$\beta_{i \ell} \mid \bm{\Lambda}, \bm Y, \bm z, \bm X  \stackrel{\text{iid}}{\sim} \text{Beta}\left(Z_{i\ell} + a, n_i - Z_{i\ell} + b\right).$$

Next, consider $\bm z \mid \bm\Lambda, \bm Y, \bm\beta, \bm X.$
First, note that
if $X_{ij\ell} \neq Y_{\lambda_{ij}\ell},$ then $z_{ij\ell} = 1.$
If instead $X_{ij\ell} = Y_{\lambda_{ij}\ell},$ then
$z_{ij\ell}\sim\text{Bernoulli}[[q_{ij\ell}/(q_{ij\ell}+\bar{q}_{i\ell})]$,
where $\bar{q}_{i\ell}=1-\beta_{i\ell}$ and
\[
q_{ij\ell}=\begin{cases}\beta_{i\ell}\,\alpha_\ell(X_{ij\ell})\,h_\ell(Y_{\lambda_{ij}\ell})\exp\left[-c\,d(X_{ij\ell},Y_{\lambda_{ij}\ell})\right]&\text{ if }\ell\le p_s,\\
\beta_{i\ell}\,\alpha_\ell(X_{ij\ell})&\text{ if }\ell>p_s.\end{cases}
\]

We now turn to the conditional distribution of $\bm Y \mid \bm\Lambda, \bz,\bb, \bX.$ Each $Y_{j'\ell}$ takes values in the set $S_\ell,$ which consists of all values for the $\ell$th field that appear anywhere in the data. This implies that
$Y_{j\prime \ell} \mid \lam, \bz, \bb, \bX$ takes the form
\[
P(Y_{j\prime \ell} = w \mid \lam, \bz, \bb, \bX) = \frac{\phi_{j'\ell}(w)}{\Phi_{j'\ell}}
\]
for all $w \in S_\ell,$ where $\Phi_{j'\ell} =\sum_{w \in S_\ell} \phi_{j'\ell}(w).$
Let $R_{j^\prime} = \{(i,j) : \lambda_{ij} = j^\prime \}$ 
be the set of all records that correspond to individual $ j^\prime.$
Immediately $\phi_{j'\ell}(w)=0$ if there exists $(i,j)\in R_{j'}$ such that $z_{ij\ell}=0$ and $X_{ij\ell}\ne w$.  If instead, for all $(i,j)\in R_{j'}$, either $z_{ij\ell}=1$ or $X_{ij\ell}=w$, then
\[
\phi_{j'\ell}(w)=\begin{cases}\alpha_\ell(w)\exp\left\{\sum_{(i,j)\in R_{j'}}z_{ij\ell}\left[\log h_\ell(w)-c\,d(X_{ij\ell},w)\right]\right\}&\text{ if }\ell\le p_s,\\ \alpha_\ell(w)&\text{ if }\ell>p_s.\end{cases}
\]

Finally, we consider the conditional distribution of  $\lam\mid\bm Y,\bz,\bb,\bX$, where 
\[
P(\lambda_{ij} = v \mid \bm Y, \bz, \bb, \bX)=\frac{\psi_{ij}(v)}{\Psi_{ij}}
\]
for all $v\in\{1,\ldots,N\}$, where $\Psi_{ij}=\sum_{v=1}^N\psi_{ij}(v)$.
Note immediately that
$\psi_{ij}(v)=0$
if there exists $\ell$ such that $z_{ij\ell} = 0$ and $X_{ij\ell} \neq Y_{v\ell}$.
If instead, for all $\ell$, either $z_{ij\ell}=1$ or $X_{ij\ell}=Y_{v\ell}$, then
\[
\psi_{ij}(v)=\exp\left\{\sum_{\ell=1}^{p_s}z_{ij\ell}\left[\log h_\ell(Y_{v\ell})-c\,d(X_{ij\ell},Y_{v\ell})\right]\right\}.
\]

\begin{rem}
The categorical fields affect the conditional distribution of $\bm\Lambda\mid\bm Y,\bz,\bb,\bX$ only insofar as they exclude certain values from the support of each distribution altogether.  If a particular field of a particular record is distorted, then it carries no information about the latent individual to which the  record should be linked.  On the other hand, if the field is not distorted, then it restricts the possible latent individuals to only those that coincide with the record in the field in question (between or among which the field conveys no preference).
\end{rem}

\section{Application to RLdata500}
\label{sec:rldata500}

%
%
%

To investigate the performance of our proposed methodology compared to existing methods, we considered the  \texttt{RLdata500} data set from the R \texttt{RecordLinkage} package, which has been considered (in some form) by \citet{steorts_2014_hash, Christen05, ChristenPudjijono09, ChristenVatsalan13}.  This simulated data set consists of 500~records, each with a first and last name and full date of birth.
These records contain 50~records that are intentionally constructed as ``duplicates'' of other records in the data set, with randomly generated errors.
The data set also includes a unique identifier for each record, so that we know we compare our methods to ``ground truth." The particular type of data found here is one in which duplication is fairly rare.

We briefly review the four classifications of how pairs of records can be linked or not linked under the truth and under the estimate.  There are four possible classifications. 
First, record pairs can be linked under both the truth and under the estimate, which we refer to as \emph{correct links} (CL). Second, record pairs can be linked under the truth but \emph{not} linked under the estimate, which are called
\emph{false negatives} (FN).
Third, record pairs can be \emph{not} linked under the truth but linked under the estimate, which are called
\emph{false positives} (FP).
Fourth and finally, record pairs can be \emph{not} linked under the truth and also \emph{not} linked under the estimate, which we refer to as \emph{correct non-links} (CNL). 
The vast majority of record pairs are classified as correct non-links in most practical settings.
Then the true number links is $\text{CL}+\text{FN}$, while the estimated number of links is $\text{CL}+\text{FP}$.  The usual definitions of false negative rate and false positive rate are
\begin{equation}\notag
\text{FNR}=\frac{\text{FN}}{\text{CL+FN}},\qquad
\text{FPR}=\frac{\text{FP}}{\text{FP+CNL}}.
\end{equation}
However, FPR as defined above is not an appropriate measure of record linkage performance, since the very large number of correct non-links (CNL) ensures that virtually any method will have an extremely small FPR, regardless of its actual quality.

Instead, we assess performance in terms of false positives by replacing FPR with the false discovery rate, i.e., the proportion of estimated links that are incorrect:
\begin{equation}\notag
\text{FDR}=\frac{\text{FP}}{\text{CL+FP}},
\end{equation}
where by convention we take $\text{FDR}=0$ if its numerator and denominator are both zero, i.e., if there are no estimated links.
Note that if the four classification pairs are laid out as a $2\times2$ contingency table, then $1-\text{FNR}$ and $1-\text{FDR}$ correspond to the number of correct links as a fraction of its row and column totals (in some order).
Thus, FDR serves as another natural counterpart to FNR.

We applied our proposed methodology to the \texttt{RLdata500} data set with $a=1$ and $b=99$, which corresponds to a prior mean of~$0.01$ for the distortion probabilities.  Also, we took $c=1$ in the string distortion distribution.
We treated birth year, birth month, and birth day as categorical variables. We treated first and last names as strings and took the string distance $d(\cdot,\cdot)$ as Edit distance \citep{winkler_2006, christen_2011}.
We ran 400,000 iterations of the Gibbs sampler described in Section~\ref{sec:gibbs}. Note that the Gibbs sampler provides a sample from the posterior distribution of the linkage structure (as well as the other parameters and latent variables).
{Note that we take the entire Gibbs sampling run as the MCMC output, i.e., we do not ``thin'' the chain or remove a ``burn-in.''}
We assess the convergence of our Gibbs sampler for the linkage structure in Figure \ref{fig:conv-rldata}. 
Furthermore, for each of the chains, the Geweke diagnostic does not reveal any immediately apparent convergence problems.

For comparison purposes, we also implemented five existing record linkage approaches for the \texttt{RLdata500} data.
Two of these methods were the simple approaches that link two records if and only if they are identical (``Exact Matching'') and that link two records if and only if they disagree on no more than one field (``Near-Twin Matching'').

The remaining three methods are regression-based procedures that treat each pair of records as a match or non-match.  Each procedure takes as covariates the Edit distance for first names and for last names, as well as the indicators of agreement on birth year, month, and day.
To reduce the number of record pairs under consideration, we first implemented a screening step that automatically treats records as non-matched if
the median of their five covariate values (i.e., their five similarity scores) is less than~$0.8$.
Hence, the remaining three methods are applied to only those record pairs that are \emph{not} excluded by the screening criterion ($99$ pairs, including all $50$ true matches).  The first regression-based method considered was the approach of Bayesian additive regression trees (BART) \citep{george_2010} with a binomial response and probit link, and with 200 trees in the sum.  Next, we applied the random forests procedure of \citet{breiman_2001} for classification, with 500 trees.  Finally, we considered ordinary logistic regression.  For each method, we fit the model on 10\%, 20\%, and 50\% of the data (i.e., the training set) and evaluated its performance on the remainder (i.e., the testing set).  For each training data percentage, we repeated the fit for 100 randomly sampled training/testing splits and calculated the overall error rate as the average of the error rates obtained by using each of the 100 splits.  We also fit and evaluated each model on the full data.

Note that we only considered methods that can take advantage of the string-valued nature of the name variables, since any method that treats these variables as categorical is unlikely to be competitive.  In particular, this rules out the approach of \citet{liseo_2011} and the SMERE procedure of \citet{steorts_2013b}.

The performance of our proposed empirical Bayesian method and the other approaches in terms of FNR and FDR is shown in Table~\ref{tab:fnr-fdr-rldata500}.  Note that by the construction of the data set, the exact matching approach produces no estimated links, so trivially its FNR and FDR are $1$ and $0$, respectively.
Since our EB method does not rely on training data, the FNR and FDR simply are what they are, which are both very low for this data set. We compare to BART, Random Forests, and logistic regression, where we reiterate that for each model we used training splits of 10\%, 20\%, and 50\%. 
{We repeated each procedure 100 times and averaged the results.}
Moreover, as is well known for supervised methods, the
apparent error rates can be reduced
when more training data is used to fit the model. For example, when we compare the EB method with the supervised methods (10\% training), our method beats each supervised procedure in both FNR and FDR. We illustrate that the error rates can be brought down if the
amount of
training data is increased, but this raises the question of whether the supervised procedure is overfitting.

{The EB method produces very low FNR and FDR compared to the supervised learning methods. We see that each supervised method is sensitive to how much training data is used, which is not desirable, and that often \emph{both} low FNR and low FDR cannot be achieved for the supervised methods.
We also}
point out that there is already an unfair advantage given to the supervised
methods
over
the
unsupervised
methods.
However, if we truly are being empirical and using the data twice, this raises the question of which method has the advantage over the other. Since these methods are not easily comparable, this also needs investigation in future work.

\begin{table}[h]
\centering
\begin{tabular}{lcc}\toprule
Procedure&FNR&FDR\\\midrule
Empirical Bayes&0.02&0.04\\
Exact Matching&1&0\\
{Near-Twin Matching}&{0.08}&0\\
BART (10\% training)&0.10&0.16\\
BART (20\% training)&0.07&0.11\\
BART (50\% training)&0.03&0.04\\
Random Forests (10\% training)&0.05&0.15\\
Random Forests (20\% training)&0.04&0.09\\
Random Forests (50\% training)&0.02&0.06\\
Logistic Regression (10\% training)&0.09&0.16\\
Logistic Regression (20\% training)&0.06&0.07\\
Logistic Regression (50\% training)&0.02&0.01\\
\bottomrule
\end{tabular}
\caption{False negative rate (FNR) and false discovery rate (FDR) for the proposed EB methodology and five other record linkage methods as applied to the \texttt{RLdata500} data.
}
\label{tab:fnr-fdr-rldata500}
\end{table}

We also calculated some additional information to assess the performance of our methodology.  The linkage structure implies a certain number of distinct individuals for the data set, which we
call~{$N_{\distinct}$. Our Gibbs sampler provides a sample from the posterior distribution of~$N_{\distinct}$, which is plotted below in Figure~\ref{fig:posterior-rldata500}.  The posterior mean is~449, while the posterior standard deviation is~7.2.   (Note that the true number of distinct individuals in the data set is~$450$.)  

\begin{figure}[ht]
\centering
\includegraphics[scale=0.55]{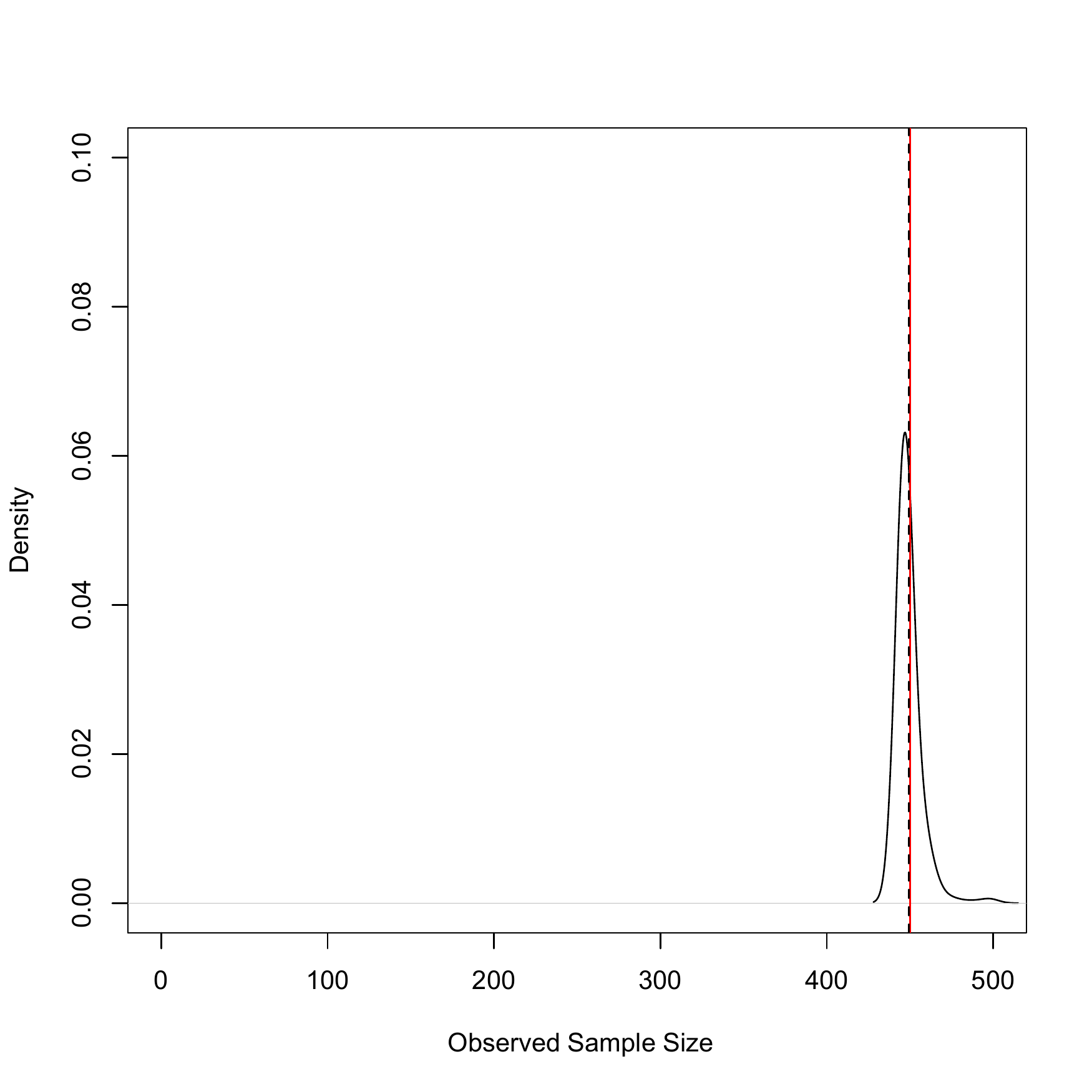}
\caption{{Posterior density of the number of distinct individuals in the sample for the \texttt{RLdata500} dataset under the proposed methodology, along with the posterior mean (black dashed line) and true value (red line).}}
\label{fig:posterior-rldata500}
\hspace{0.5cm}
\end{figure}

\begin{figure}[htbp]
\begin{center}
\includegraphics[width=\textwidth]{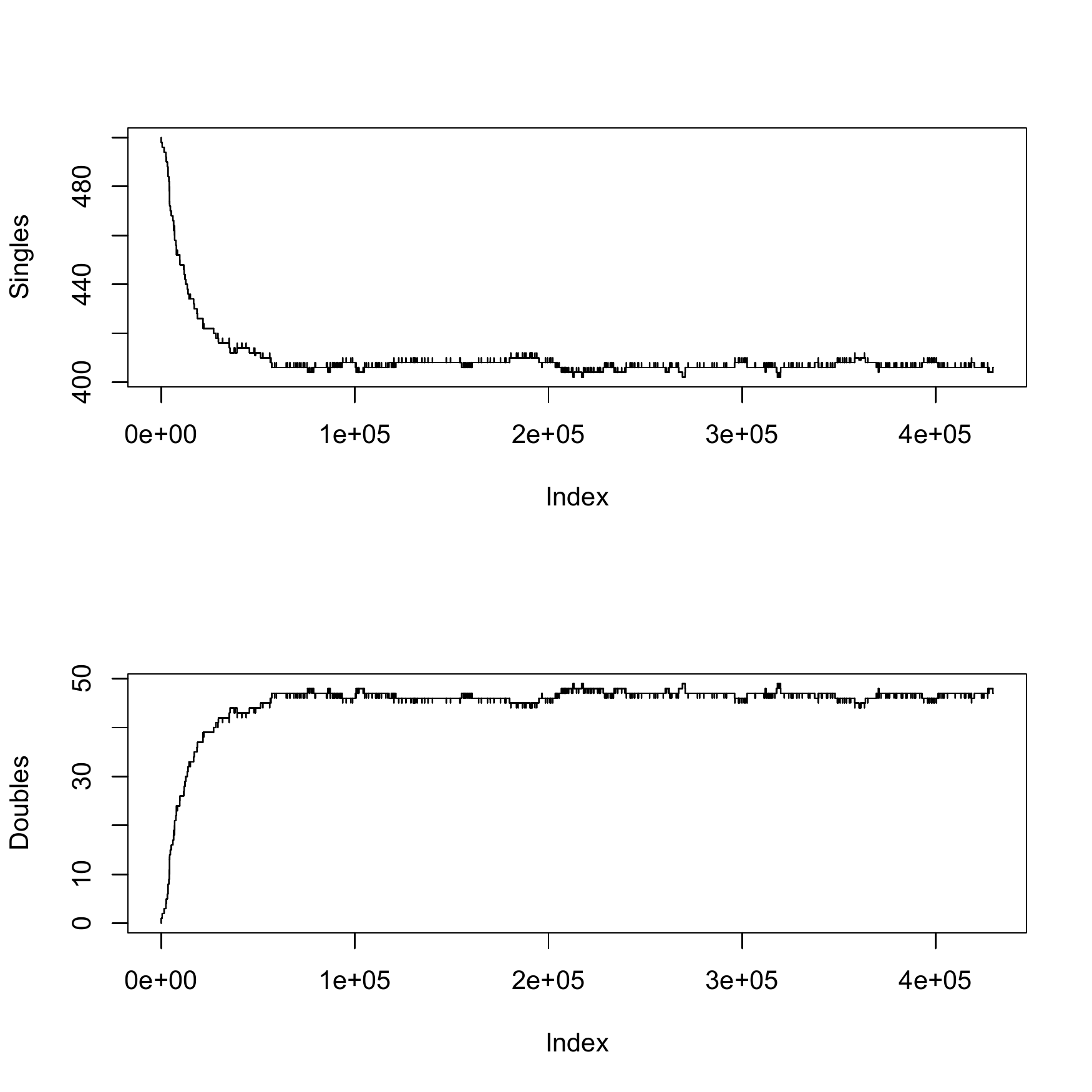} 
\caption{Trace plots of the number of latent entities that are represented in the sample by exactly one record (``singles'') and by exactly two records (``doubles'') for 400,000 Gibbs samples for the \texttt{RLdata500} dataset.}
\label{fig:conv-rldata}
\end{center}
\end{figure}

\section{Application to Italian Household Survey}
\label{sec:italy}
We also evaluated the performance of our proposed methodology using data from the Italian Survey on Household and Wealth (FWIW), a sample survey conducted by the Bank of Italy every two years. The 2010 survey covered 19,836 individuals, while the 2008 survey covered  19,907 individuals.  
The goal is to merge the 2008 and 2010 lists
by considering the following categorical variables: year of birth, working status, employment status, branch of activity, town size, geographical area of birth, sex, whether or not Italian national, and highest educational level obtained.
Note in particular that data about individuals' names is \emph{not} available, which makes record linkage on this data set a challenging problem.  (However, a unique identifier \emph{is} available to serve as the ``truth.'')  
As in Section~\ref{sec:rldata500}, we evaluate performance using false negative rate (FNR) and false discovery rate (FDR). 

We applied our proposed methodology to a subset of this data (region 6; all other regions exhibit similar behavior) with $a=1$ and $b=99$, which corresponds to a prior mean of $0.01$ for the distortion probabilities. 
Also, we took $c=1$ in the string distortion distribution.  We treated all variables here as categorical. 
We ran 10,000 iterations of the Gibbs sampler described in Section~\ref{sec:gibbs}, which took approximately 10~hours.

In principle, we would also apply the same methods as in Section~\ref{sec:rldata500} (BART, random forests, and logistic regression).  These methods essentially treat each pair of records as an observation.  Since the number of record pairs is very large (242,556 record pairs arising from 697 observations from region 6), it is necessary to first reduce the number of record pairs under consideration by a screening rule to eliminate pairs that are clearly non-linked.  For the data of Section~\ref{sec:rldata500}, it was straightforward to find a screening rule (based on the median of the similarity scores) that greatly reduced the number of record pairs under consideration while still including all pairs that were truly linked.  However, we could not find any viable screening rule for this data, at least in part because all fields are categorical.  More specifically, any screening rule of the form ``eliminate a record pair unless it agrees on at least $K$ out of a particular set of $M$ fields'' either inadvertently eliminates some true links or retains far too many record pairs (at least 44,426).  In practice, of course, the elimination of some true links is not a major problem, as it simply creates some automatic false negatives.  However, the application of such a screening method is inappropriate if the goal is to evaluate the performance of a record linkage method, since the automatic false negatives would create a substantial handicap that is not the fault of the method itself.  (Still, the necessity of such a screening method \emph{is} an inherent disadvantage of any method that treats each record pair as an observation.  Of course, our proposed empirical Bayesian model does not suffer from this problem.)

Since it is not clear how to obtain a fair comparison of our methodology to BART, random forests, or logistic regression,
we instead compare to other methods:
the approach of \citet{liseo_2011} and the SMERE approach of \citet{steorts_2013b}.
We also compare to the approaches of exact and ``near-twin" matching.
The approach of \citet{liseo_2011} took 3 hours, while SMERE took 20~minutes. Under the recommendation of \cite{liseo_2011}, we ran 100,000 iterations of the Gibbs sampler, which we also did for SMERE. 


Turning to convergence of the
{Gibbs sampler for our method,}
we again look at trace plots as we did for the \texttt{RLdata500} data as can be seen in Figure \ref{fig:convergence}. 
Based on these plots, it appears fairly safe to treat the MCMC sample as an approximate draws from the posterior distribution (not necessarily independent, however).  Thus,
Table~\ref{table:fnr-fdr-italy} compares the FNR and FDR of our proposed EB methodology to that of the approach of \citet{liseo_2011} and of SMERED from \cite{steorts_2013b}.
{We note that SMERED and the EB method perform about the same, and vastly improved upon the method of \cite{liseo_2011}.}
Again, we reiterate that this data set consists solely of categorical variables that provide relatively little information by which to link or separate records, hence, the large error rates in Table~\ref{table:fnr-fdr-italy} are not surprising. We note that the number of links missed among twins and near-twins is 28,246, so any method will do poorly on this type of data without a field attribute that helps the linkage procedure drastically. This is shown very well by the FNR and FDR in Figure~\ref{fig:posterior-italy}. 
Again, this is not a weakness of the method, but of the feature-poor data.

\begin{table}[htbp]
\centering
\begin{tabular}{lcc}\toprule
Procedure&FNR&FDR\\ \midrule
Empirical Bayes&0.34&0.36\\
Tancredi--Liseo&0.52&0.46\\
SMERE &0.33&0.29\\
{Exact Matching} & {0.29} & {0.70} \\
{Near-Twin Matching} & {0.14} & {0.98} \\
\bottomrule
\end{tabular}
\caption{False negative rate (FNR) and false discovery rate (FDR) for the proposed empirical Bayesian methodology and two other record linkage methods as applied to the Italian Household Survey data.}
\label{table:fnr-fdr-italy}
\end{table}

As in Section~\ref{sec:rldata500}, we again examined the posterior distribution of the number of distinct individuals in the data set.  This posterior, which has mean 498.8 and standard deviation 0.48, is shown in  Figure~\ref{fig:posterior-italy}. (We provide a sensitivity analysis in Section~\ref{sec:sensitivity}.) 

\begin{figure}[htbp]
\begin{center}
\includegraphics[width=0.5\textwidth]{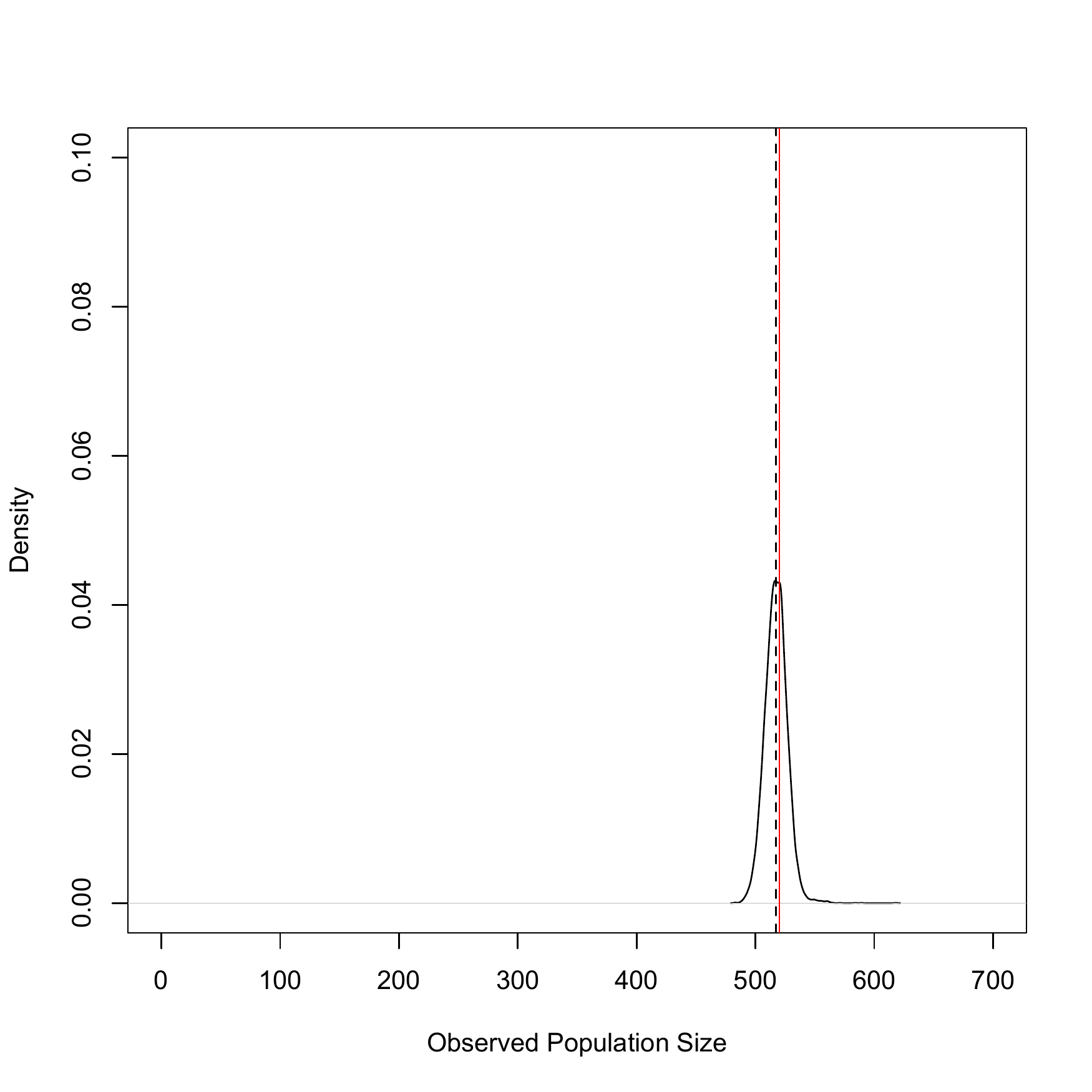}
\caption{{Posterior density of the number of distinct individuals in the sample for the Italian data under the proposed EB-type methodology, along with the posterior mean (black dashed line) and true value (red line).}}
\label{fig:posterior-italy}
\end{center}
\end{figure}

\begin{figure}[htbp]
\begin{center}
\includegraphics[width=0.65\textwidth]{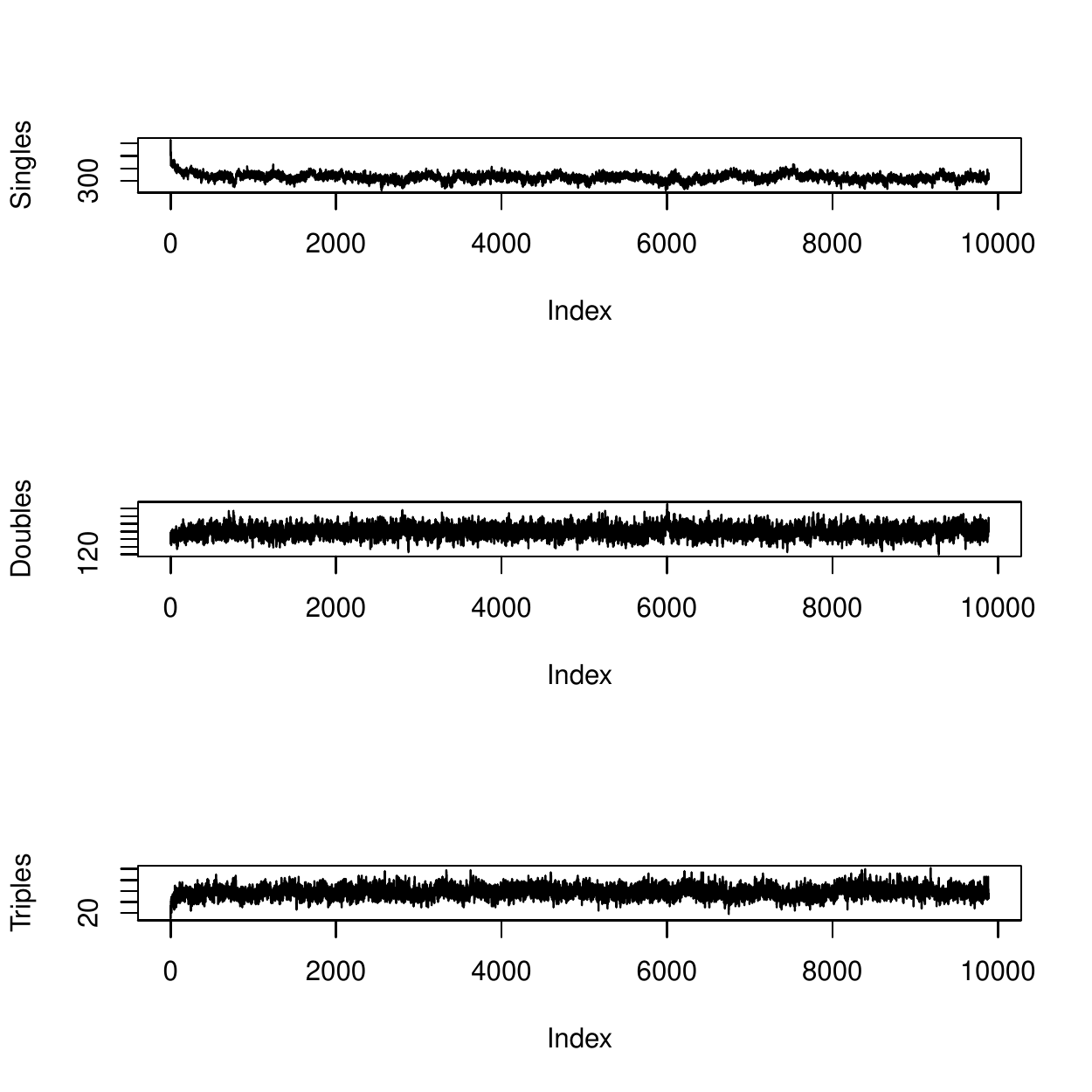}
\caption{Trace plots of the number of latent entities that are represented in the sample bye exactly one record (``singles''), by exactly two records (``doubles''), and by exactly three records (``triples'') for 10,000 Gibbs samples of the Italian dataset. }
\label{fig:convergence}
\end{center}
\end{figure}

\section{Robustness to Prior Specification}
\label{sec:sensitivity}

In Sections \ref{sec:rldata500} and~\ref{sec:italy}, we investigated the performance of our proposed methodology on data sets from RLdata500 and the FWIW.  To do so, we made specific choices for various quantities in the model in~(\ref{model:eb}).  In particular, we chose values of the hyperparameters $a$ and $b$ that determine the prior for the distortion probabilities.  We also chose a steepness parameter~$c$ and a string metric~$d$ to govern the distortion distributions of string-valued fields.  Finally, we chose a value of the effective latent population size~$N_{\pop}$.  In practice, however, it may not be immediately clear how to make these choices when faced with an unfamiliar application or data set.  Hence, it is of interest to know how robust the model in~(\ref{model:eb}) is to changes in these various quantities.

\paragraph{{RLdata500 data}} We begin with the \texttt{RLdata500} data.  For each Gibbs sampling run described below, {we executed 100,000 iterations.}

We first consider the effect of varying the values of $a$ and $b$, while fixing $c=1$ and $N_{\pop}=500$ with $d$ as edit distance.  Note that the prior distribution of the distortion probabilities is $\text{Beta}(a,b)$, so $a/(a+b)$ is the prior mean for these distortion probabilities.  Moreover, for any fixed value of $a/(a+b)$, increasing the values of $a$ and~$b$ proportionally decreases the variance of this prior distribution.  Figure~\ref{fig:run1} shows the results obtained by fixing $a/(a+b)=0.002$ and varying $a$ and $b$ proportionally.  It can be seen from the left-most posterior densities that when $b<10,$ the posterior underestimates the truth. We also see this behavior more clearly by looking at how the posterior mean and posterior standard deviation change as we vary $a,b$ (see Table \ref{tab:run1}).

We also consider the effect of varying the ratio $a/(a+b)$ while holding $a+b$ fixed at either $a+b=100$ (top plot of Figure~\ref{fig:run2}) or $a+b=10$ (bottom plot of Figure~\ref{fig:run2})), with $c$, $d$, and $N_{\pop}$ the same as in Figure~\ref{fig:run1}). 
 We see that in the top plot when we vary the prior mean and when this value is high (10 percent), this causes over-linkage and the observed sampled size is too low. In the bottom figure, we find the bottom three in the legend are clumping in the same place around 390--400 for the observed sample size. The only one that is close to ground truth is $a=0.0003, b=9.9997.$ 
We find from this plot (as in Figure \ref{fig:run1}) that when $b< 10,$ 
the model tends to underestimate the observed sample size. This makes sense because
the value of $b$ in a Beta distribution controls how fast the distribution dies off for larger probabilities. Thus, setting $b$ too small makes it very likely that you will have distortion probabilities that are moderate. Both the behavior just described of both plots is reinforced by 
Tables~\ref{tab:run2} and \ref{tab:run22}.

Next, we vary the choice of~$c$, the steepness parameter of the string distortion distribution.  Note that the larger the value of~$c$, the less likely it is for string-valued record fields to be distorted to values that are far (as measured by the string metric~$d$) from their corresponding latent entity's field value.  We set $a=0.01$, $b=$99, and $N_{\pop}=$500, and we took $d$ to be~edit distance.  The results are shown in Figure~\ref{fig:c}.  We see that resulting estimated posterior is sensitive to the choice {of~$c.$}

We also considered two different string metrics, the aforementioned edit distance as well as Jaro-Winkler distance~\citep{winkler_2006}, for use as the distance~$d$ in the string distortion distribution.  We set $c=1$ and $N_{\pop}=500$, and we took a few choices for $a$ and~$b$ for each string metric.  {The result for Jaro-Winkler distance is shown in Figure~\ref{fig:edit-and-jw}.  (The corresponding plot for edit distance is shown in the aforementioned Figure~\ref{fig:run1}.)}  We see that when $b<10$ under both choices of $d,$ the estimated posterior is greatly underestimated. We see that as $b$ increases and $a$ is quite small, then the posterior is more concentrated around the true posterior and true observed sample size (red line).
We see this same behavior in Tables~\ref{tab:run1} and \ref{tab:jw}.

Finally, we investigate the effect of different choices of the effective latent population size~$N_{\pop}$.  We consider values of $N_{\pop}$ both smaller and larger than the sample size, while fixing $a=$0.01, $b=$99, and $c=1$ and taking $d$ to be edit distance.  The results are shown in Figure~\ref{fig:M}.  We see that
when we use $N_{\pop}= 450$ and $N_{\pop}= 550,$ we find the posterior means are 448 and 457, and the posterior standard deviations are 1 and 7 respectively (running the chain for 30,000 iterations). When we use $N_{\pop} = 1000$, we find the posterior mean is 479 with a posterior standard deviation of 10. 
It can be seen that small changes to $N_{\pop}$ do not yield dramatic differences in the posterior distribution of the observed sample size, but larger changes to $N_{\pop}$ can have a more substantial effect.  Thus, determination of a procedure or guideline for choosing an appropriate value of $N_{\pop}$ is an important goal of future study.

\begin{figure}[htbp]
\begin{center}
\includegraphics[width=0.45\textwidth]{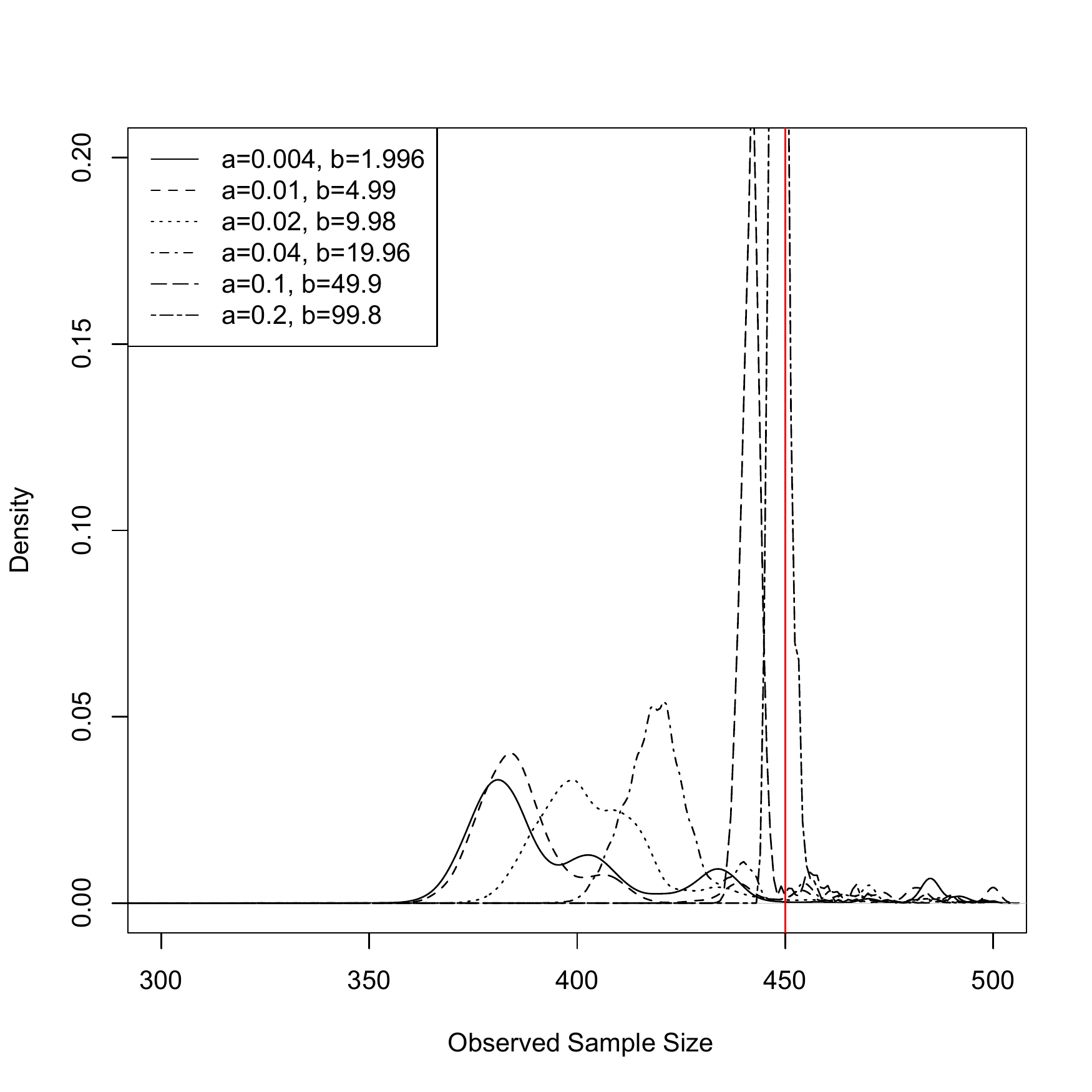}
\caption{{Posterior density of the number of distinct individuals in the sample for the \texttt{RLdata500} data for several values of $a$ and~$b$.  Note that $a/(a+b)=0.002$ in all cases.  The red line marks the true value.}}
 \label{fig:run1}
\end{center}
\end{figure}

\begin{table}[htbp]
\begin{center}
\begin{tabular}{cc|cc}
$a$ & $b$ & posterior mean  & standard deviation \\ \hline
0.004 & 1.996 & 398.35 & 28.45  \\ 
0.010 & 4.990 & 398.58 & 31.15 \\
0.020 & 9.980 & 407.07 & 19.09\\
0.040 & 19.96 & 422.67 & 13.69 \\
0.100 & 49.90 & 442.78 &5.71 \\
0.200 & 99.80 &  447.37 &   6.20 \\
\end{tabular}
\end{center}
\caption{{Posterior mean and standard deviation of the number of distinct individuals in the sample for the \texttt{RLdata500} data for several values of $a$ and~$b$ (compare to Figure~\ref{fig:run1}).}}
\label{tab:run1}
\end{table}%

\begin{figure}[htbp]
\begin{center}
\includegraphics[width=0.65\textwidth]{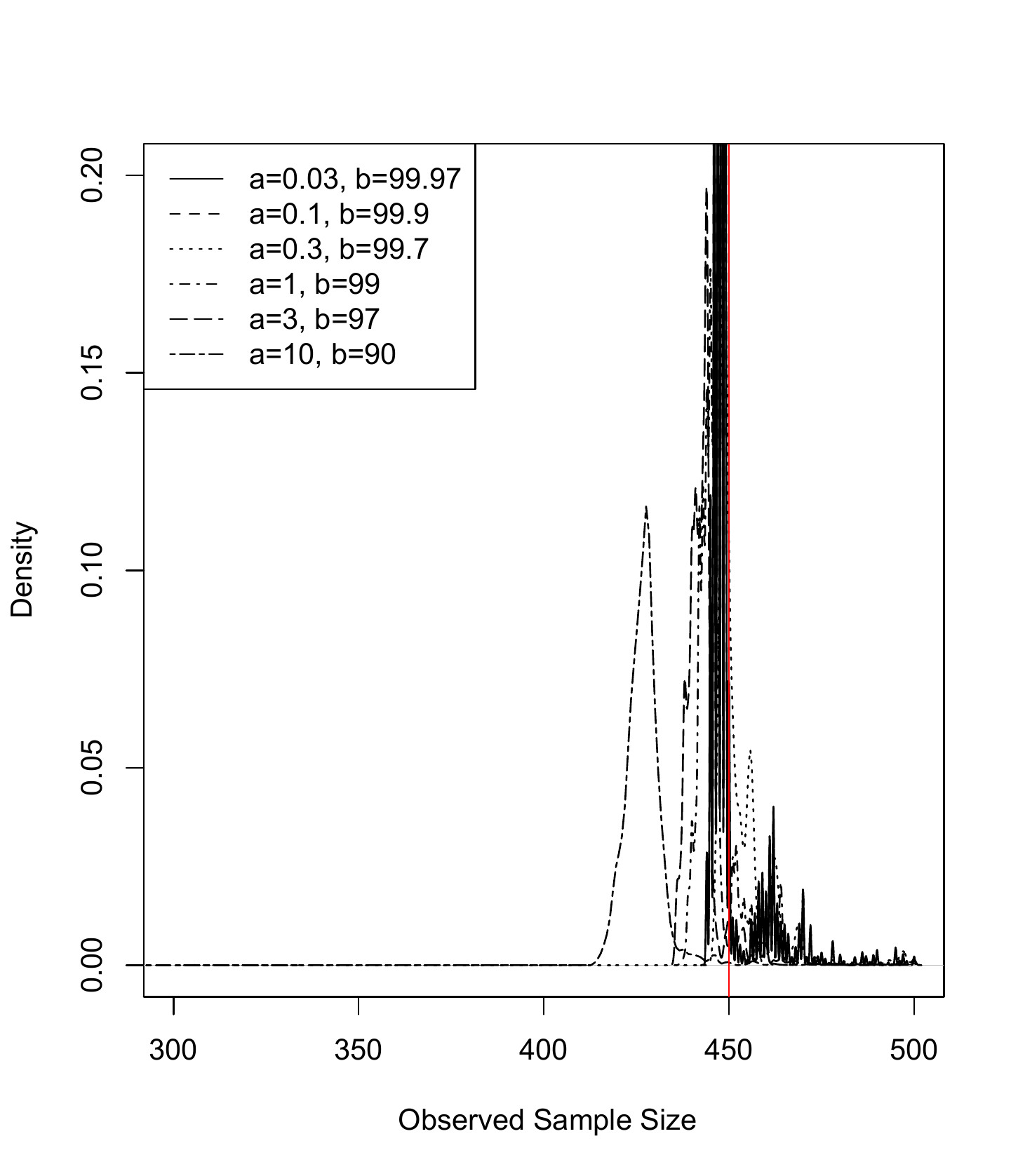}
\includegraphics[width=0.65\textwidth]{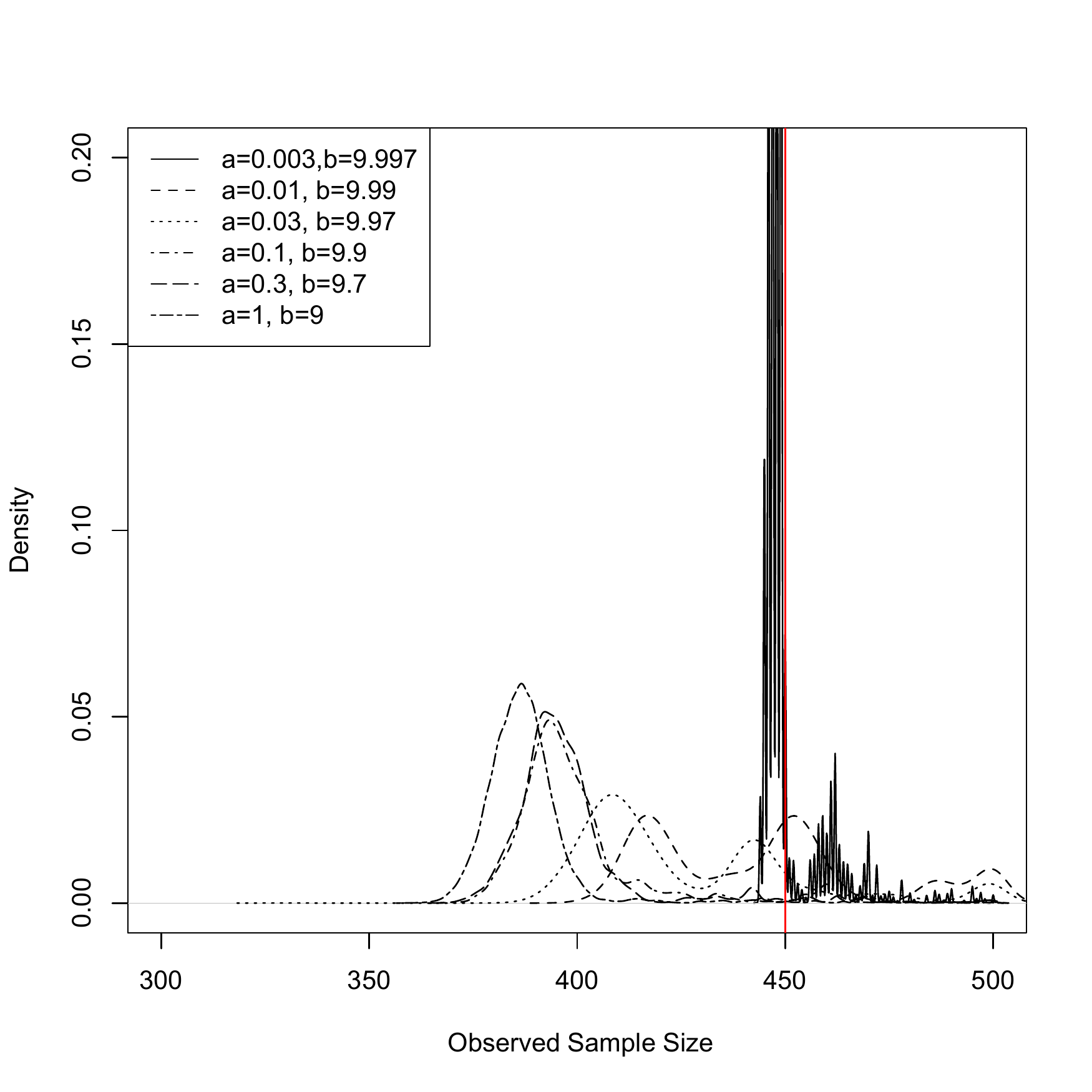}
\caption{{Posterior density of the number of distinct individuals in the sample for the \texttt{RLdata500} data for several values of $a$ and~$b$.  The top plot fixes $a+b=100$ in all cases, while the bottom plot fixes $a+b=10$ in all cases.  The red line marks the true value.}}
\label{fig:run2}
\label{default}
\end{center}
\end{figure}

\begin{table}[htbp]
\begin{center}
\begin{tabular}{cc|cc}
$a$ & $b$ & posterior mean  & standard deviation \\ \hline
0.03 & 99.97 &452.6889 & 10.32819 \\ 
0.1 & 99.99 & 447.0832 & 4.862139 \\ 
0.3 & 99.97 & 447.2618 & 4.900142 \\ 
1 & 99 & 445.3222 &  4.098847 \\ 
3 & 97 & 441.6043 & 3.611562 \\ 
10 & 90 & 426.177 & 4.607082 \\
\end{tabular}
\end{center}
\caption{{Posterior mean and standard deviation of the number of distinct individuals in the sample for the \texttt{RLdata500} data for several values of $a$ and~$b$ (compare to the top plot of Figure~\ref{fig:run2}).}}
\label{tab:run2}
\end{table}%

\begin{table}[htbp]
\begin{center}
\begin{tabular}{cc|cc}
a & b & posterior mean  & standard deviation \\ \hline
0.003 & 9.997 &430.7297& 27.22453 \\ 
0.01 & 9.99 & 410.5785 & 21.98859 \\ 
0.03 & 9.97 & 403.055 & 17.24064 \\ 
0.1 & 9.9 & 395.769 &  11.26418\\ 
0.3 & 9.7 & 392.6065 & 9.743864 \\ 
1 & 9 & 386.1448 & 8.609332 \\
\end{tabular}
\end{center}
\caption{{Posterior mean and standard deviation of the number of distinct individuals in the sample for the \texttt{RLdata500} data for several values of $a$ and~$b$ (compare to the bottom plot of Figure~\ref{fig:run2}).}}
\label{tab:run22}
\end{table}%

\begin{figure}[htbp]
\begin{center}
\includegraphics[width=0.65\textwidth]{final_plots/run5_finalplot_edit}
\includegraphics[width=0.65\textwidth]{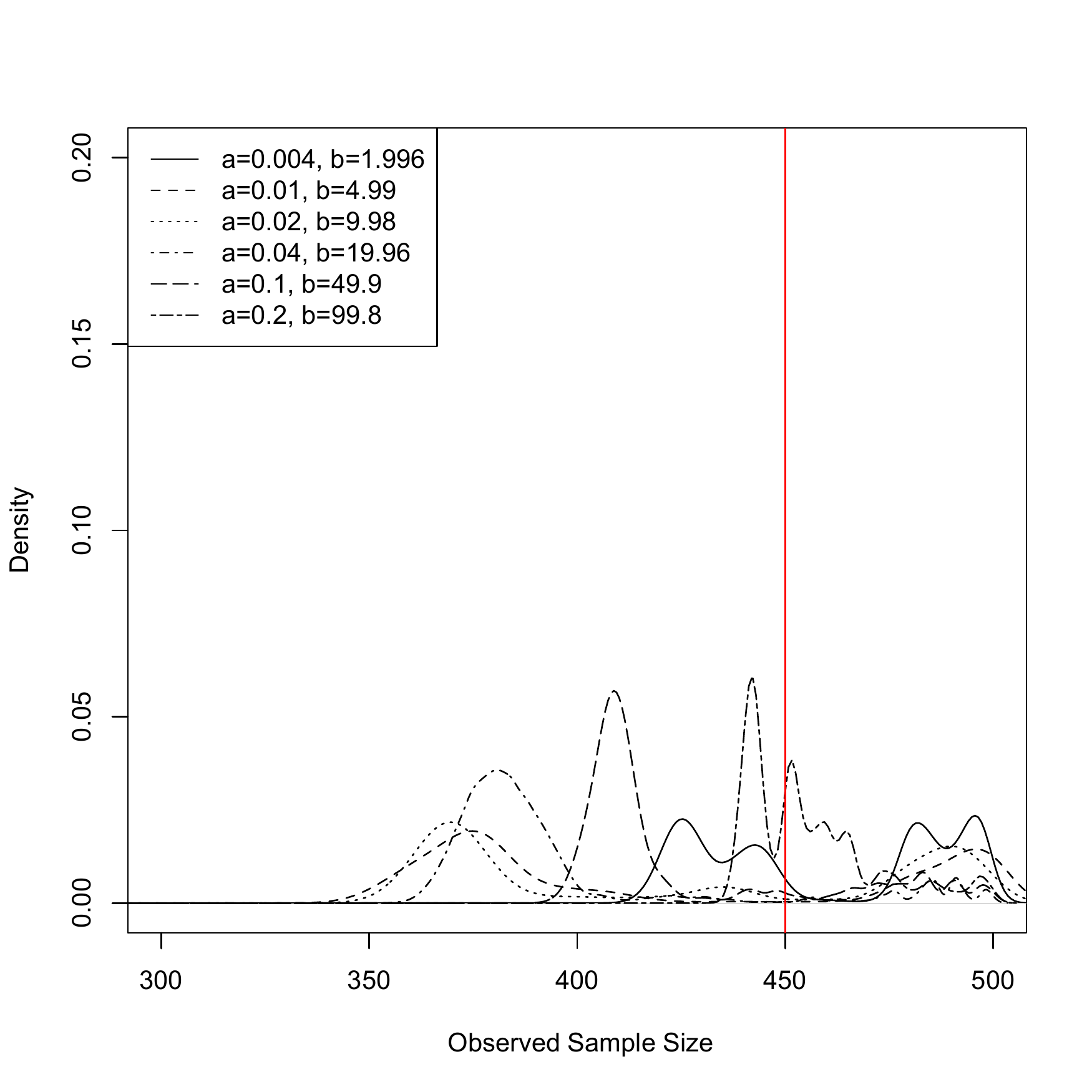}
\caption{{Posterior density of the number of distinct individuals in the sample for the \texttt{RLdata500} data for several values of $a$ and~$b$ using Jaro-Winkler distance instead of edit distance in the string distortion distribution.  Note that $a/(a+b)=0.002$ in all cases.  The red line marks the true value.}}
\label{fig:edit-and-jw}
\end{center}
\end{figure}

\begin{table}[htbp]
\begin{center}
\begin{tabular}{cc|cc}
$a$ & $b$ & posterior mean   & standard deviation \\ \hline
0.004 & 1.996 & 459.49425 &29.04779  \\ 
0.010 & 4.990 & 417.7381& 56.48478 \\
0.020 & 9.980 & 421.621 & 54.85565\\
0.040 & 19.96 & 395.0226 & 33.55952 \\
0.100 & 49.90 & 424.9393 & 29.36313 \\
0.200 & 99.80 &  455.7053 &   15.59177 \\
\end{tabular}
\end{center}
\caption{{Posterior mean and standard deviation of the number of distinct individuals in the sample for the \texttt{RLdata500} data for several values of $a$ and~$b$ using Jaro-Winkler distance instead of edit distance in the string distortion distribution (compare to Figure~\ref{fig:edit-and-jw}).}}
\label{tab:jw}
\end{table}%

\begin{figure}[ht]
\begin{center}
\includegraphics[scale=0.45]{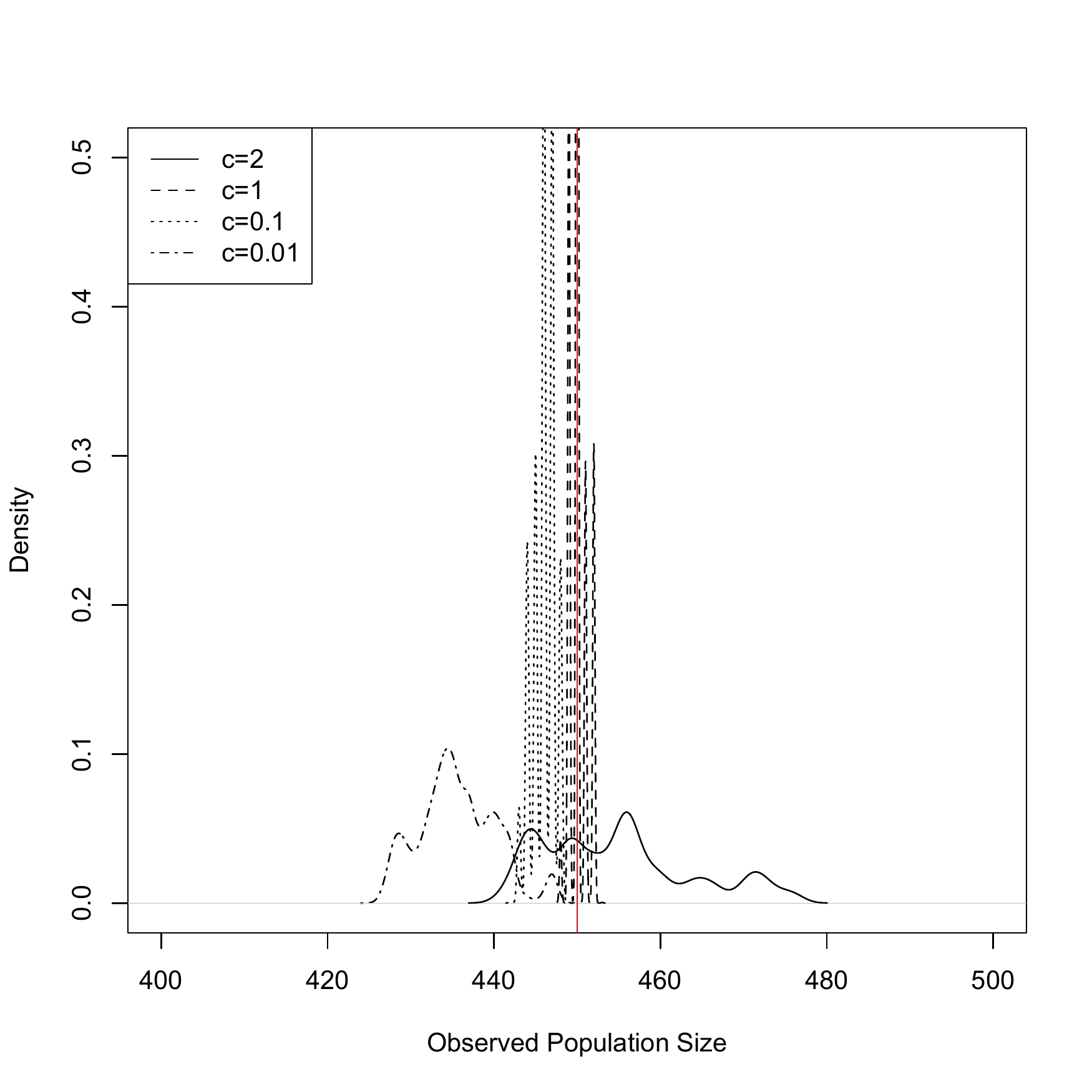}
\caption{{Posterior density of the number of distinct individuals in the sample for the \texttt{RLdata500} data for several values of~$c$, with $a=0.01$ and $b=99$.  The red line marks the true value.}}
\label{fig:c}
\end{center}
\end{figure}

\begin{figure}[htbp]
\begin{center}
\includegraphics[width=0.45\textwidth]{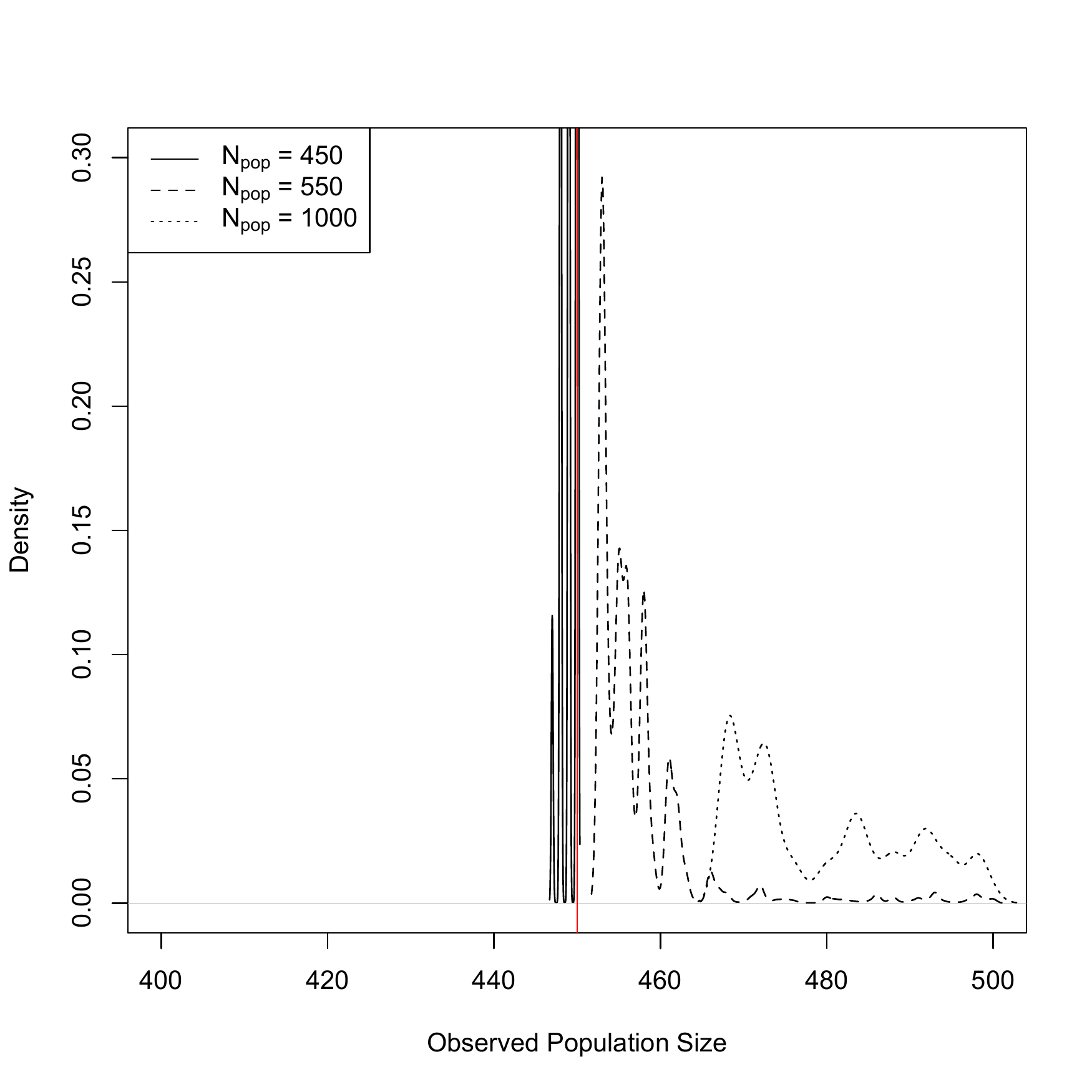}
\caption{{Posterior density of the number of distinct individuals in the sample for the \texttt{RLdata500} data for several values of the latent population size~$N_{\pop}$, with $a=0.01$ and $b=99$.  The red line marks the true value.}}
\label{fig:M}
\end{center}
\end{figure}

\clearpage
\newpage

\paragraph{Italian data}
{We also investigated the sensitivity of the Italian data results to changes in the various subjective parameters.}

{We first varied the latent population size~$N_{\pop}$ while taking $a=1$, $b=99$, $c=1$, and $d$ as edit distance.  Each Gibbs sampling run consisted of 30,000 iterations.  The results are shown in Figure~\ref{fig:italy-varyM} and Table~\ref{tab:italian_varyM}, and they are broadly similar to those observed for the \texttt{RLdata500} data discussed previously.}

{We also considered various values of $a$ and~$b$, with $c=1$, $N_{\pop}=1300$, and edit distance used as the distance metric in the string distortion distribution.  Each Gibbs sampling run consisted of 30,000 iterations.  We began by varying $a$ and~$b$ both together with their ratio held constant, as shown in Figure~\ref{fig:italian:run1} and Table~\ref{tab:italian:run1}.  Next, we varied $a$ and~$b$ separately with their sum held constant, as shown in Figure~\ref{fig:italian:run2} and Tables~\ref{tab:italian:run2}~and~\ref{tab:italian:run22}.  Again, the results were fairly similar to those for the \texttt{RLdata500} data.}

\begin{figure}[htbp]
\begin{center}
\includegraphics[width=0.45\textwidth]{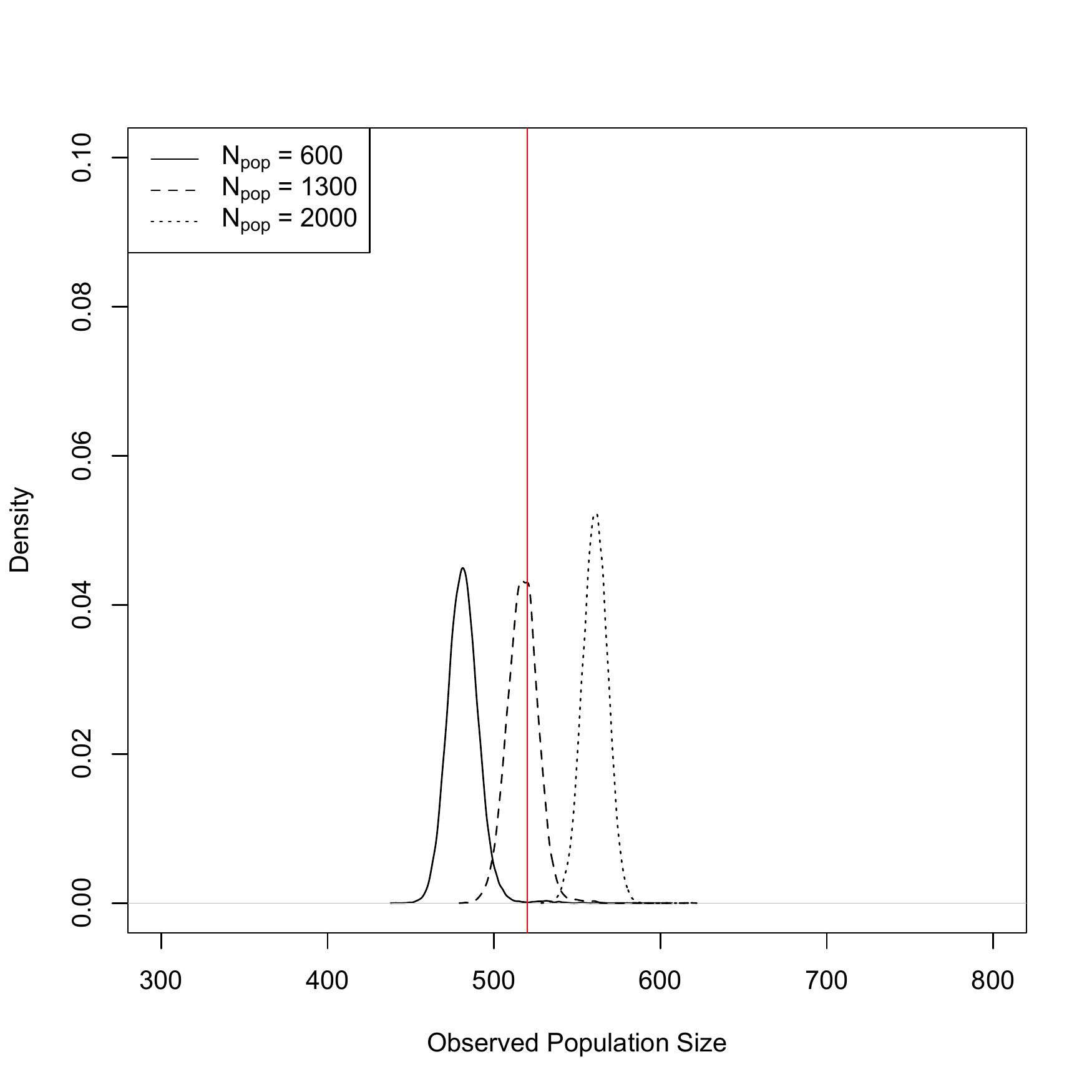}
\caption{{Posterior density of the number of distinct individuals in the sample for the Italian data for several values of the latent population size~$N_{\pop}$, with $a=1$ and $b=99$.  The red line marks the true value.}}
\label{fig:italy-varyM}
\end{center}
\end{figure}

\begin{table}[htbp]
\begin{center}
\begin{tabular}{c|cc}
$N_{\pop}$  & posterior mean  & standard deviation \\ \hline
600 & 400.65 & 7.95 \\
1300 & 517.35 & 9.42 \\
2000 & 560.614 & 7.89 \\
\end{tabular}
\end{center}
\caption{{Posterior mean and standard deviation of the number of distinct individuals in the sample for the Italian data for several values of $N_{\pop}$ (compare to Figure~\ref{fig:italy-varyM}).}}
\label{tab:italian_varyM}
\end{table}%

\begin{figure}[htbp]
\begin{center}
\includegraphics[width=0.45\textwidth]{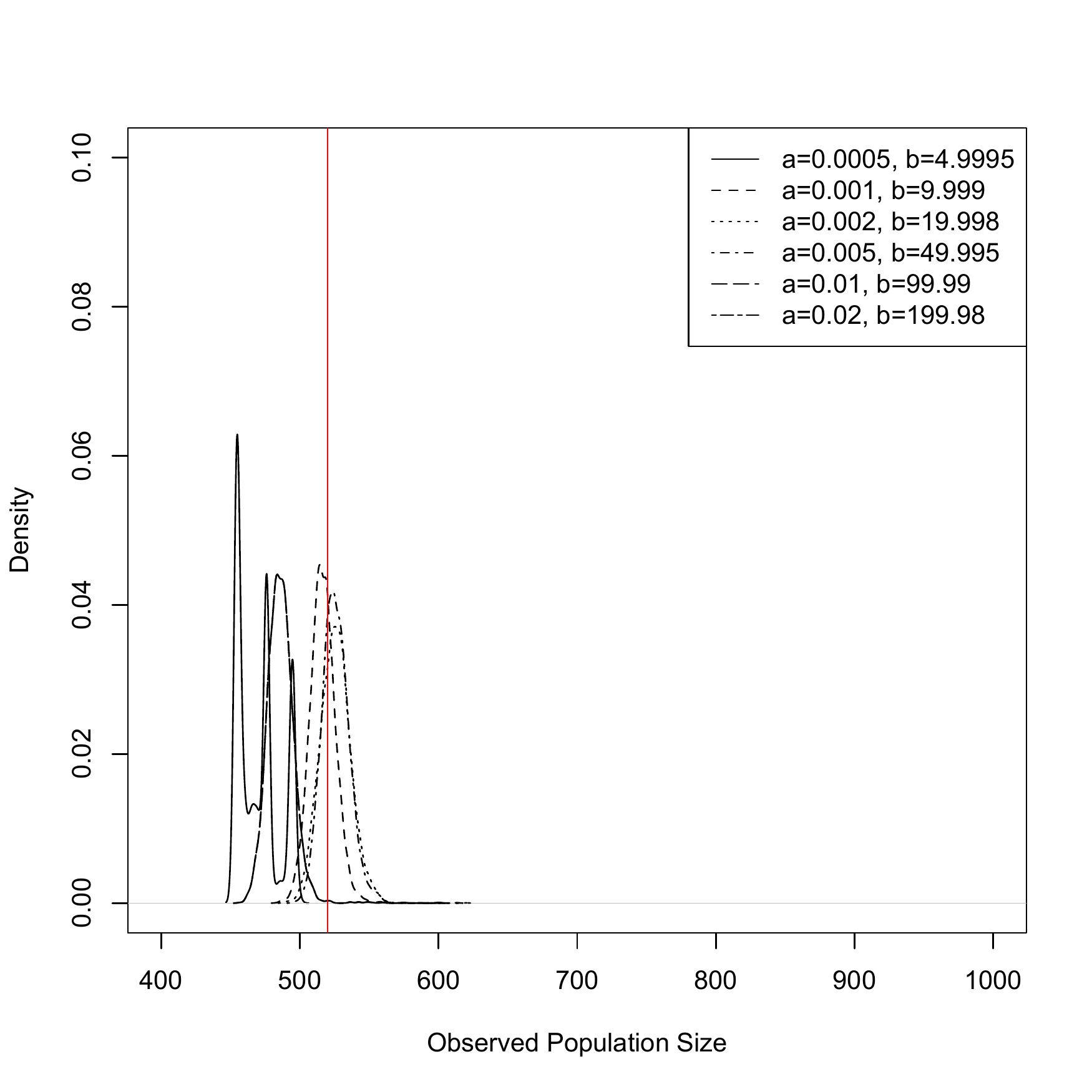}
\caption{{Posterior density of the number of distinct individuals in the sample for the Italian data for several values of $a$ and $b$.  Note that $a/(a+b)=0.0001$.  The red line marks the true value.}}
\label{fig:italian:run1}
\end{center}
\end{figure}

\begin{table}[htbp]
\begin{center}
\begin{tabular}{cc|cc}
$a$ & $b$ & posterior mean   & standard deviation \\ \hline
0.0005 & 4.9995 & 470.2311 & 14.87979 \\
0.001 & 9.999 & 516.5542 & 9.333234 \\ 
0.002 & 19.998 & 525.6803 & 10.94388 \\
0.005 & 49.995 & 525.8361 & 9.770544 \\
0.01 & 99.99 & 486.0217 & 9.669656 \\ 
 0.02 & 199.98 & 486.0217 & 9.669656  \\
\end{tabular}
\end{center}
\caption{{Posterior mean and standard deviation of the number of distinct individuals in the sample for the Italian data for several values of $a$ and~$b$ (compare to Figure~\ref{fig:italian:run1}).}}
\label{tab:italian:run1}
\end{table}%

\begin{figure}[htbp]
\begin{center}
\includegraphics[width=0.45\textwidth]{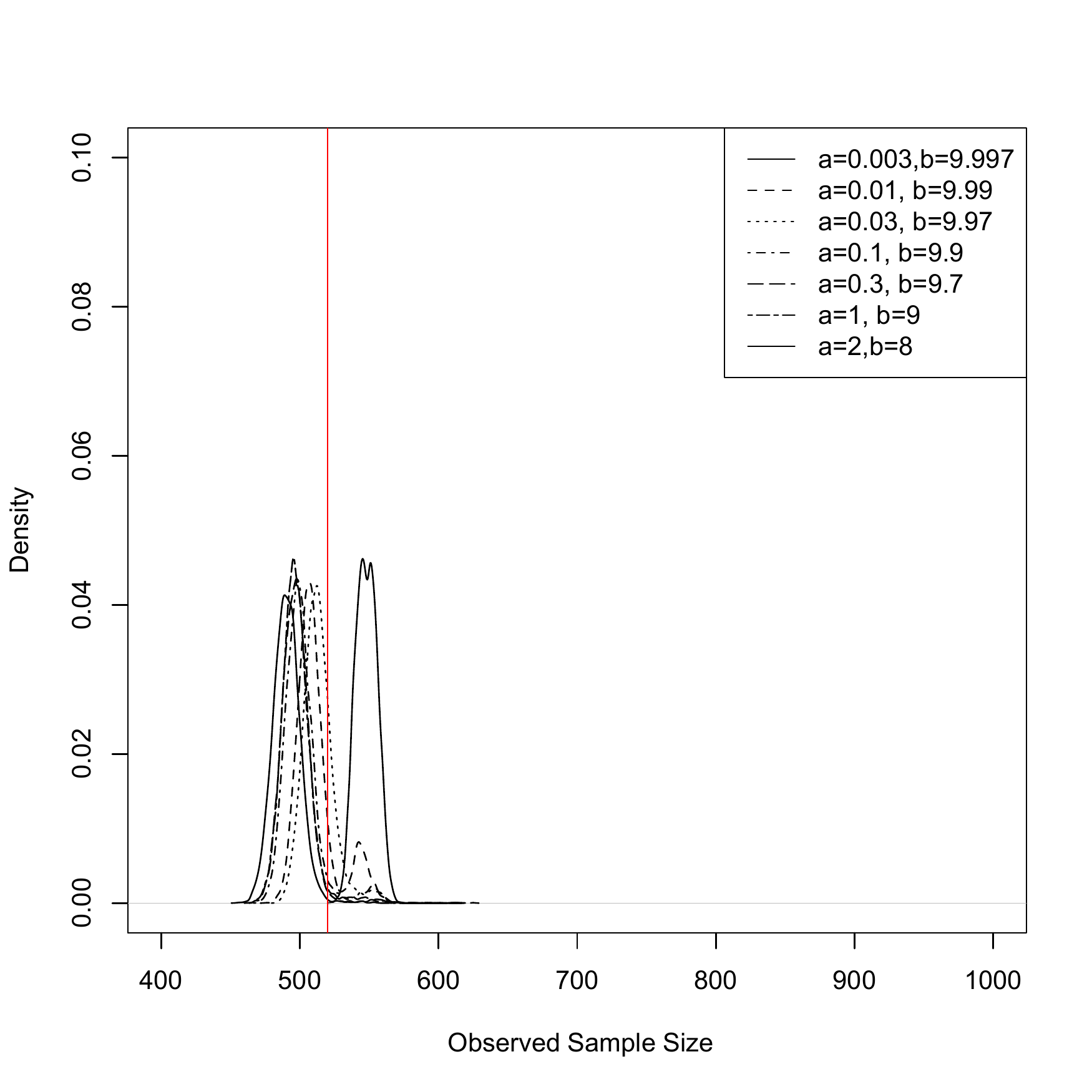}
\includegraphics[width=0.45\textwidth]{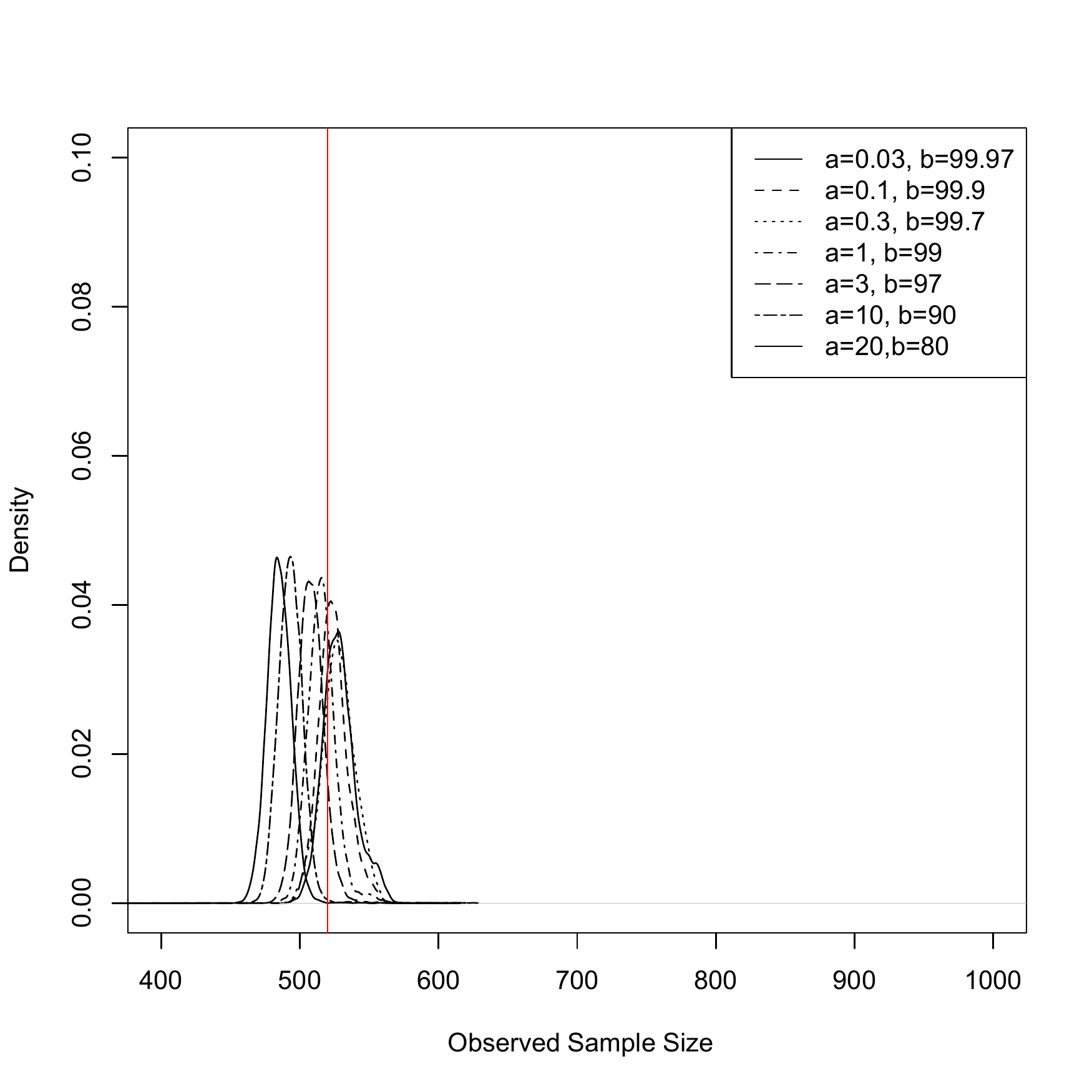}
\caption{{Posterior density of the number of distinct individuals in the sample for the Italian data for several values of $a$ and $b$.  The left plot fixes $a+b=10$ in all cases, while the right plot fixes $a+b=100$ in all cases.  The red line marks the true value.}}
\label{fig:italian:run2}
\end{center}
\end{figure}

\begin{table}[htbp]
\begin{center}
\begin{tabular}{cc|cc}
$a$ & $b$ & posterior mean  & standard deviation \\ \hline
0.03 & 99.97 &528.5115 & 12.02456 \\ 
0.1 & 99.99 &524.0653 & 11.87278 \\ 
0.3 & 99.97 & 527.7922 & 11.86949 \\ 
1 & 99 & 507.8796&10.123 \\ 
3 & 97 & 493.3781 & 9.100615\\ 
10 & 90 & 485.0756 & 9.199441 \\
\end{tabular}
\end{center}
\caption{{Posterior mean and standard deviation of the number of distinct individuals in the sample for the Italian data for several values of $a$ and~$b$ (compare to the right plot of Figure~\ref{fig:italian:run2}).}}
\label{tab:italian:run2}
\end{table}%

\begin{table}[htbp]
\begin{center}
\begin{tabular}{cc|cc}
$a$ & $b$ & posterior mean  & standard deviation \\ \hline
0.003 & 9.997 & 554.4813 &  5.389025\\ 
0.01 & 9.99 & 528.9712 & 19.40173 \\ 
0.03 & 9.97 & 521.1844 & 15.62619 \\ 
0.1 & 9.9 &510.0594 &  20.46328\\ 
0.3 & 9.7 &  504.7957 & 16.15656 \\ 
1 & 9 & 500.853 & 17.39763\\
2 & 8 & 494.7102 &  13.51739
\\
\end{tabular}
\end{center}
\caption{{Posterior mean and standard deviation of the number of distinct individuals in the sample for the Italian data for several values of $a$ and~$b$ (compare to the left plot of Figure~\ref{fig:italian:run2}).}}
\label{tab:italian:run22}
\end{table}%


\section{Discussion}
We have made
several
main contributions with this paper. 
First, we have extended the categorical record linkage and de-duplication methodology of \citet{steorts_2013b}
to a new approach that handles both categorical and string-valued data, while using the same linkage structure~$\bm{\Lambda}$.
This extension to string-valued data makes our approach flexible enough to accommodate a variety of applications.
Note that all of the various benefits of the approach of \citet{steorts_2013b} are obtained by our new formulation.
In particular, the ability to calculate posterior matching probabilities leads to \emph{exact} error propagation (as opposed to merely providing bounds) when estimates arising from the record linkage model are subsequently integrated into other types of analyses (e.g., capture-recapture techniques for estimating population size).
Moreover, our proposed empirical Bayesian approach retains the aforementioned benefits of the Bayesian paradigm while eliminating the need to specify subjective priors for the latent individuals.
{Indeed, the \emph{only} subjective parameters that must be specified at all are the values $a$ and $b$ that determine the distribution of the distortion probabilities, the value~$c$ that appears in the string distortion distribution, and the latent population size~$N_{\pop}$.}
We demonstrated our method by applying it to a simulated data set for which accurate record linkage is fairly easy and a real data set for which accurate record linkage is quite difficult.
We found that our method compares favorably to a collection of popular supervised learning methods and another standard Bayesian method in the literature.

Our work serves as 
an early entry into the literature of empirical Bayesian record linkage methodology, and it can likely be improved, extended, and tailored to fit particular problems and applications.
We believe that unsupervised methods,
such as our proposed method,
have a clear advantage over supervised approaches since in most applications, training data is scarce or unavailable altogether and in many cases the validity of the training data cannot be checked or trusted. 


It is clear from both the present work and the results of \citet{steorts_2013b} that Markov chain Monte Carlo (MCMC) procedures impose serious computational limitations on the database sizes that are addressable by these Bayesian record linkage techniques.  Since real record linkage applications often involve databases with millions of records, there is the possibility that MCMC-based Bayesian inference may not be the most promising direction for future research. Possible solutions may be provided by the variational Bayesian literature.
Variational approximations work by systematically ignoring
some dependencies among the variables being inferred, bounding the error this
introduces into the posterior distribution, and minimizing the bound.  If properly
chosen, the minimization is a fast optimization problem and the minimal error
is small. Such techniques have long been used to
allow Bayesian methods to scale to
industrial-sized data sets in domains such as topic modeling \citep{wainwright_2008, broderick_2013}.  
In particular, the framework developed by \citet{wainwright_2008, broderick_2013}  allows for a full posterior distribution.
This is
appealing for record linkage methodology
since it would allow quick estimation of posterior matching probabilities for propagation into subsequent analyses.
It is also possible that the computational difficulties of the Bayesian record linkage approach could be circumvented by some other altogether different approach, such as the formulation of a model for which various posterior quantities of interest are calculable in closed form or via more manageable numerical procedures.


\subsubsection*{Acknowledgements}
We would like to thank the excellent comments and suggestions from 
the referees, associate editor, and editor that have led to major improvements
of this paper. RCS was supported by National Science Foundation through
grants SES1130706 and DMS1043903 and the John Templeton Foundation.
All views expressed in this work are of the author alone and not the grants/foundations.


\newpage
\clearpage
\bibliography{chomp}
\bibliographystyle{ba}

\newpage
\clearpage

\section*{Appendix}
\label{sec:appendix}

\section*{Joint Posterior Derivation}
We derive the joint posterior below.
\label{app:joint}
\begin{align*}
&\pi(\bm\lambda,\bm Y,\bm z,\bm\beta\mid\bm X)\\
&\propto
\left[\prod_{i=1}^k\prod_{j=1}^{n_i}
\left(\prod_{\ell=1}^{p_s}\left\{(1-z_{ij\ell})\,I(X_{ij\ell}=Y_{\lambda_{ij}\ell})+\frac{z_{ij\ell}\,
\textcolor{black}{
\alpha_\ell(X_{ij\ell})\,
}
\exp\!\left[-c\,d(X_{ij\ell},Y_{\lambda_{ij}\ell})\right]}{\sum_{w\in S_\ell}
\textcolor{black}{
\alpha_\ell(w)\,
}
\exp\!\left[-c\,d(w,Y_{\lambda_{ij}\ell})\right]}\right\}\right.\right.\\
&\qquad\qquad\qquad\left.\left.\times\prod_{\ell=p_s+1}^{p_s+p_c}\left\{(1-z_{ij\ell})\,I(X_{ij\ell}=Y_{\lambda_{ij}\ell})+z_{ij\ell}\,\alpha_\ell(X_{ij\ell})\right\}\right)\right]\\
&\qquad\times\left[\prod_{j'=1}^N\prod_{\ell=1}^{p_s+p_c}\alpha_\ell(Y_{j'\ell})\right]\left[\prod_{i=1}^k\prod_{j=1}^{n_i}\prod_{\ell=1}^{p_s+p_c}\beta_{i\ell}^{z_{ij\ell}}(1-\beta_{i\ell})^{1-z_{ij\ell}}\right]\\
&\qquad\times\left[\prod_{i=1}^k\prod_{\ell=1}^{p_s+p_c}\beta_{i\ell}^{a-1}(1-\beta_{i\ell})^{b-1}\right]\left[\prod_{i=1}^k\prod_{j=1}^{n_i}I(\lambda_{ij}\in\{1,\ldots,N\})\right]\\
&\propto
\left[\prod_{i=1}^k\prod_{j=1}^{n_i}
\left(\prod_{\ell=1}^{p_s}\left\{(1-z_{ij\ell})\,I(X_{ij\ell}=Y_{\lambda_{ij}\ell})+\frac{z_{ij\ell}\,
\textcolor{black}{
\alpha_\ell(X_{ij\ell})\,
}
\exp\!\left[-c\,d(X_{ij\ell},Y_{\lambda_{ij}\ell})\right]}{\sum_{w\in S_\ell}
\textcolor{black}{
\alpha_\ell(w)\,
}
\exp\!\left[-c\,d(w,Y_{\lambda_{ij}\ell})\right]}\right\}\right.\right.\\
&\qquad\qquad\qquad\left.\left.\times\prod_{\ell=p_s+1}^{p_s+p_c}\left\{(1-z_{ij\ell})\,I(X_{ij\ell}=Y_{\lambda_{ij}\ell})+z_{ij\ell}\,\alpha_\ell(X_{ij\ell})\right\}\right)\right]\\
&\qquad\times\left[\prod_{j'=1}^N\prod_{\ell=1}^{p_s+p_c}\alpha_\ell(Y_{j'\ell})\right]\left[\prod_{i=1}^k\prod_{\ell=1}^{p_s+p_c}\beta_{i\ell}^{\sum_{j=1}^{n_i}z_{ij\ell}+a-1}(1-\beta_{i\ell})^{n_i-\sum_{j=1}^{n_i}z_{ij\ell}+b-1}\right]\\
&\qquad\times\left[\prod_{i=1}^k\prod_{j=1}^{n_i}I(\lambda_{ij}\in\{1,\ldots,N\})\right]\\
&\propto
\prod_{i=1}^k\prod_{j=1}^{n_i}\left(
\left\{\mathop{\prod_{\ell=1}^{p_s+p_c}}_{z_{ij\ell}=0}I(X_{ij\ell}=Y_{\lambda_{ij}\ell})\right\}\right.\\
&\qquad\qquad\qquad\left.\times\left\{\mathop{\prod_{\ell=1}^{p_s}}_{z_{ij\ell}=1}\frac{
\textcolor{black}{
\alpha_\ell(X_{ij\ell})\,
}
\exp\!\left[-c\,d(X_{ij\ell},Y_{\lambda_{ij}\ell})\right]}{\sum_{w\in S_\ell}
\textcolor{black}{
\alpha_\ell(w)\,
}
\exp\!\left[-c\,d(w,Y_{\lambda_{ij}\ell})\right]}\right\}
\left\{\mathop{\prod_{\ell=p_s+1}^{p_s+p_c}}_{z_{ij\ell}=1}\alpha_\ell(X_{ij\ell})\right\}\right)\\
&\qquad\times\left[\prod_{j'=1}^N\prod_{\ell=1}^{p_s+p_c}\alpha_\ell(Y_{j'\ell})\right]\left[\prod_{i=1}^k\prod_{\ell=1}^{p_s+p_c}\beta_{i\ell}^{\sum_{j=1}^{n_i}z_{ij\ell}+a-1}(1-\beta_{i\ell})^{n_i-\sum_{j=1}^{n_i}z_{ij\ell}+b-1}\right]\\
&\qquad\times\left[\prod_{i=1}^k\prod_{j=1}^{n_i}I(\lambda_{ij}\in\{1,\ldots,N\})\right].
\end{align*}
\textcolor{black}{
If we 
restrict the allowed values of $\lambda_{ij}$ to the set $\{1,\ldots,N\}$, then the last line above is irrelevant.  Also, for each $w_0\in S_\ell$, define the quantity
$
h_\ell(w_0)=\left\{\sum_{w\in S_\ell}\exp\!\left[-c\,d(w,w_0)\right]\right\}^{-1},
$
i.e., $h_\ell(w_0)$ is the normalizing constant for the distribution
$F_\ell(w_0)$.  We can compute $h_\ell(w_0)$ in advance for each possible
$w_0\in S_\ell$. We can simplify the posterior to
\begin{align*}
&\pi(\bm\lambda,\bm Y,\bm z,\bm\beta\mid\bm X)\\
&\propto
\prod_{i=1}^k\prod_{j=1}^{n_i}\left(
\left\{\mathop{\prod_{\ell=1}^{p_s+p_c}}_{z_{ij\ell}=0}I(X_{ij\ell}=Y_{\lambda_{ij}\ell})\right\}\left\{\mathop{\prod_{\ell=1}^{p_s+p_c}}_{z_{ij\ell}=1}\alpha_\ell(X_{ij\ell})\right\}\left\{\mathop{\prod_{\ell=1}^{p_s}}_{z_{ij\ell}=1}h_\ell(Y_{\lambda_{ij}\ell})\right\}\right.\\
&\qquad\qquad\qquad\left.\times\exp\!\left[-c
\sum_{\ell=1}^{p_s}z_{ij\ell}\,
d(X_{ij\ell},Y_{\lambda_{ij}\ell})\right]
\right)\\
&\qquad\times\left[\prod_{j'=1}^N\prod_{\ell=1}^{p_s+p_c}\alpha_\ell(Y_{j'\ell})\right]\left[\prod_{i=1}^k\prod_{\ell=1}^{p_s+p_c}\beta_{i\ell}^{\sum_{j=1}^{n_i}z_{ij\ell}+a-1}(1-\beta_{i\ell})^{n_i-\sum_{j=1}^{n_i}z_{ij\ell}+b-1}\right]\\
&\propto
\prod_{i=1}^k\prod_{j=1}^{n_i}\left\{
\left[\mathop{\prod_{\ell=1}^{p_s+p_c}}_{z_{ij\ell}=1}\alpha_\ell(X_{ij\ell})\right]\left[\mathop{\prod_{\ell=1}^{p_s}}_{z_{ij\ell}=1}h_\ell(Y_{\lambda_{ij}\ell})\right]\exp\!\left[-c
\sum_{\ell=1}^{p_s}z_{ij\ell}\,
d(X_{ij\ell},Y_{\lambda_{ij}\ell})\right]
\right\}\\
&\qquad\times\left[\prod_{j'=1}^N\prod_{\ell=1}^{p_s+p_c}\alpha_\ell(Y_{j'\ell})\right]\left[\prod_{i=1}^k\prod_{\ell=1}^{p_s+p_c}\beta_{i\ell}^{\sum_{j=1}^{n_i}z_{ij\ell}+a-1}(1-\beta_{i\ell})^{n_i-\sum_{j=1}^{n_i}z_{ij\ell}+b-1}\right]\\
&\qquad\times I(X_{ij\ell}=Y_{\lambda_{ij}\ell}\text{ for all }i,j,\ell\text{ such that }z_{ij\ell}=0).
\end{align*}}

\section*{Comparison of SMERED and EB Model}
We compare the main differences between SMERED and the EB model proposed here. First, SMERED
assumes categorical (non-string or non-text data). Furthermore, it assumes a generative hierarchal Bayesian model, instead of an empirical Bayesian model. An HB model works very well for categorical data, however, becomes quickly intractable for noisy ``text" data. Finally, the models are similar in that both cluster records to a hypothesized latent entity. The HB model assumes the latent entity is from a Multinomial distribution, whereas, 
we assume the latent entity is drawn from the empirical distribution of either some data related to the data at hand or the data itself. This allows for faster updating of the latent quantities.

\end{document}